\documentclass[preprint,12pt]{elsarticle}
\usepackage{graphicx, color,soul, tikz, pgfplots, eurosym}
\pgfplotsset{compat=1.15}
\usetikzlibrary{backgrounds,shapes,arrows,positioning,calc,decorations.pathmorphing,fit}
\usepgflibrary{decorations.markings}
\usepackage{amssymb,amsthm, float,textcomp  }
\usepackage{amsmath}
\usepackage{subcaption}
\date{}
\makeatletter
\def\ps@pprintTitle{%
 \let\@oddhead\@empty
 \let\@evenhead\@empty
 \def\@oddfoot{\centerline{\thepage}}%
 \let\@evenfoot\@oddfoot}
\makeatother

\begin{document}

\begin{frontmatter}


\title{Resilient expansion planning of virtual energy plant with an integrated energy system- reliability criteria of lines and towers}



\author[ttu,dtu]{Sambeet Mishra\corref{mycorrespondingauthor}}
\author[ttu,elr]{Henri Manninen}
\author[dtu]{Qiuwei Wu}
\author[uit]{Chiara Bordin\corref{mycorrespondingauthor}}

\cortext[mycorrespondingauthor]{Corresponding author}
\address[ttu]{Tallinn University of Technology, Estonia}
\address[dtu]{Danish Technical University, Denmark}
\address[uit]{UiT, The Arctic University of Norway, Norway}
\address[elr]{Elering, Estonian electricity system operator, Estonia}

\begin{abstract}
Virtual power plants, while being virtual, rely on a physical network for operations. The portfolio of virtual power plant is flexible in facilitating a wide range of resources including the local heat pumps. The power transmission network has the responsibility to ensure security of supply, reliability of operation, planning and expansion. The power transmission network and apparatus including lines and towers are also ageing with time. Furthermore, the transmission network covers a large geographical area which is expensive to maintain.

The objective of this paper is to investigate the effect of power network condition on the power network expansion planning. The condition of power network is determined by maintenance cost of lines and a health index and a risk factor associated to tower. The investigation begins with answering how the inclusion of heat pump impacts the decisions on network interventions. Thereafter, the condition network is factored in into the decision making by better understanding the impact of the network condition on the overall expansion planning. Furthermore many sensitivity analysis are conducted to evaluate the trade-offs between decision variables such as cost of heat pump, coefficient of performance of heat pump, risk factors and line and tower costs. 
 
\end{abstract}


\end{frontmatter}

\begin{table}[htbp]
\label{tab:indexes}
\scriptsize
\begin{tabular}{l p{9cm}}
\bfseries Nomenclature & \\
\hline
\textbf{Indexes}\\
\hline
$t$ & Time step\\
$a$ & Years \\
$i,j$ & Nodes of the grid\\
\hline
\bfseries Parameters & \\
\hline
$D_{i,t,a}$ & Power demand\\
$D^{heat}_{i,t,a}$ & Heat demand\\
$r$ & Interest rate\\
$E_{i,j}$ & Binary parameters equal to 1 if a line exists between node i j\\
$\overline E_{i,j}$ & Capacity of the cable\\
$Z$ & Construction time \\
$BigM$ & A very big number\\
$N^{pot}_{i,j}$ & Parameter that is equal to 1 if a potential arc can be placed between node i and j\\
$\overline N_c$ & Capacity of the new cable of type c\\
$X_{i,j}$ & Binary parameter equal to 1 if the users wants to evaluate replacement of an existing cable, 0 otherwise\\
$V^{health}_{i}$ & State of health of a tower in a nosw\\
$V^{CRF}$ & capital recovery factor for the installation of a new tower\\
$V^{life}$ & estimated life of a new tower\\
$V^{cost}$ & estimated cost of a new tower\\
$V^{risk}_{i,a}$ & risk factor associated to a node\\
$H^{CRF}$ & capital recovery factor for the installation of a new heat pump\\
$H^{life}$ & estimated life of a new heat pump\\
$H^{cost}$ & estimated cost of a new heat pump\\
$COP$ & coefficient of performance of a heat pump\\

\hline
\bfseries Variables & \\
\hline
$f^{in}_{i,t,a}$ & Total power flow into a node \\
$p_{i,j,t,a}$ & Power flow in each arc \\
$k_{i,j,c,a}$& Binary variable equal to 1 if an existing arc i,j is replaced by a potential cable of type c, 0 otherwise\\
$\gamma_{i}$ & Binary variable equal to 1 if an existing pole is replaced with a new one, 0 otherwise\\
$y_{i,j,c,a}$ & Binary variable equal to 1 if a potential arc is created between nodes i and j, 0 otherwise\\
$d^{heat01}_{i,t,a}$ & portion of heat demand to be met through existing traditional heating systems\\
$d^{heat02}_{i,t,a}$ & portion of heat demand to be met through new heat pump installation\\
$h_{i,a}$ & new aggregated heat pump capacity installed on a node i in a certain year a\\
\hline
\end{tabular}
\end{table}

\section{Introduction}
\label{S:1}

The United Nations projects that the total carbon emission need to be reduced by 7.6\% annually for a coming decade to meet the Paris climate goal to keep the temperature below 1.5\textdegree centigrade \cite{christensen2019emissions, hohne2020emissions}. In the power sector, a switch from conventional fossil fuels to renewable energy resources is rapidly underway. This massive transition comes with challenges including planning and operation of the modern power and energy sector. As several studies indicate, the sector coupling, especially integrating the electricity and energy sector, will lead to overall optimal energy systems and energy flows \cite{fridgen2020holistic}. An integrated energy system (IES) has the potential to provide not only flexibility in the overall system but also minimization of losses by efficient utilization of resources.
Renewable energy resources such as solar and wind are variable, uncertain and often distributed. To mitigate this challenge massive energy storage solutions are needed. Many large scale projects are currently underway and various energy storage technologies are being explored. In parallel, behind the meter localized solutions and demand side flexibility are emerging among the key solutions to maintain the supply-demand balance. A virtual power plant (VPP) enables pooling together decentralized and distributed generation, storage, and consumption into one platform. Thereby a VPP forms a localized and centralized energy solution which can access the energy and electricity market. Authors in \cite{chen2021} have proposed a hierarchical control strategy for VPP management for multiple grid support services. An optimal aggregation approach for VPP considering network reconfiguration is presented by the authors \cite{Zhou2021}.
It is clear from the available literature that a VPP with an integrated energy system has the potential to provide cost-effective and efficient solutions for total system planning and operations. 

The power grid connects the bulk of generation to consumption while maintaining the overall system balance and stability. The power network is continuously expanding as the demand grows and new sectors are being connected. The network also facilitates moving the distributed energy generation from the point of generation to the consumption. The system operator therefore must ensure the system stability, balance and security. The ageing network requires maintenance and supervision to function on optimal capacity. Many studies have been conducted to address how the system operator can ensure resilient power supply as in \cite{khoo2020demand, kopsidas2017power}. However the condition of the network, specifically the condition of transmission towers and overhead lines, are often not among the decision variables in the models proposed in the existing literature. In fact, many expansion planning models are built on a top-down format where the line capacities are aggregated forming a representative network. This overlooks the practical condition of a particular network which may have an adverse effect in real world performance. The authors in \cite{rao2012studies} elaborates on transmission tower failures under different scenarios. An investment planned without a proper insight on the condition of the network might mount total or partial economic losses due to operational failures in the  integrated energy system. 

In order to fill the gap in the existing literature, this paper addresses how resilient power network expansion can be performed within an IES, including the overall conditions of power network.   

VPPs are among the widely adopted technologies for a distributed, decentralized and decarbonized power network \cite{bhuiyan2021towards}. An integrated energy system acts as a physical layer of interconnection between electric and heat systems and flows. A VPP is the virtual layer on top of the physical layer that provides optimal management of assets in the overall network and market operations. A case study from Sweden demonstrates how a VPP can balance the variable renewable power generation \cite{monie2021residential}. However, very few studies address the VPP concept together with an integrated energy system to investigate how this coupling can contribute to mitigate the costly interventions in network expansion and reconfiguration within the VPP.

\subsection{Reasoning and challenges}

The system operator plans for network expansions at regular intervals to ensure an optimal system operation and a proper integration of new generation units or demand within the network. Similarly an asset owner performs regular maintenance to ensure availability and performance of the generation asset. When planning new investments, the network operator need to decide how to allocate the funds between network maintenance and capacity building. The network maintenance can be of two types: (a) restructuring - changing the structure of an existing network through dismantling of existing lines and replacing with new ones, and scheduling maintenance of a line or a corridor that consists of multiple lines (b) reconfiguration - new line capacity building to facilitate new demand or generation unit \cite{bordin2021multihorizon}. While network planning is a long term and multi-horizon decision, the state of the network is dynamically changing in short time horizon. Therefore addressing the multi-horizon network expansion in view of the dynamic deterioration of network condition is the main objective of this work. In literature, this consideration is either absent or not addressed in the context of optimal expansion planning to the best of the knowledge of the authors. 

An online remaining useful life estimation of power lines for predictive maintenance is presented in \cite{riba2021line}. A condition monitoring method for wind turbines is discussed in \cite{ren2021novel}. An algorithm to facilitate localized energy exchanges through a VPP is presented in \cite{chen2021}. In \cite{chen2020decentralized} authors present a decentralized provision algorithm to predict the renewable power production.
A resilient system planning has the objective to ensure the stability, balance, quality, safety and security of supply with demand. The scope of a VPP with an integrated energy system is widely studied in the literature \cite{zhao2020active, ye2020operation, kong2020robust, li2021novel}. Authors in \cite{navidi2021agent} propose a joint expansion planning model with ownership sharing of the energy system and transmission grid. Author in \cite{manninen2021health} present a health index prediction method for overhead transmission lines. Automated utility pole condition monitoring is presented in \cite{ramlal2020toward}. An automatic condition assessment of high-voltage transmission lines using deep learning techniques is presented in \cite{manninen2021toward}. An optimal aggregation approach for VPP considering network reconfiguration is presented by the authors \cite{Zhou2021}.

As the literature suggests, predictive and condition based maintenance are being widely adopted by the system operators and asset owners to schedule the maintenance. Therefore, the
condition of the network need to be factored in the expansion planning decision. In \cite{mishra2019rnr, bordin2021multihorizon} two mathematical optimization approaches to the reliability oriented expansion planning problem considering the restructuring and reconfiguration decisions are presented. 

This paper expands the reliability oriented expansion planning model with inclusion of towers and lines, health index and risk factor within the mathematical optimization model, with a focus on VPP coupled with an IES.   

The problem is formulated from the network operator's perspective with perfect information from a VPP. The VPP includes assets such as: heat pump, non-dispatchable generation unit based on renewable resource with battery banks, dispatchable generation units. The energy consumption is divided into electric and heat demand. 
The following sub sections introduce specific challenges addressed in this work.

\subsubsection{The challenge of heat pump capacity allocation}
Heat pump is considered as a reliable and cost-optimal solution in mitigating the local heating energy demand \cite{ruhnau2019time}. The coefficient of performance (COP) of a heat pump varies based on the underlying technology in a heat pump \cite{chabot2020numerical}. Authors in \cite{ostergaard2021variable} studied how the variable energy tariff levied on the production can promote flexibility from heat pump. The variation of COP reflects accordingly in the cost, specifically higher COP comes with higher price. Determining optimal capacity installation of a heat pump depends on several factors such as the COP of the heat pump, the cost of the heat pump, the condition of the network, etc. 
A relevant question that arise is, what is an optimal ratio of heat pump capacity to COP to meet the heat energy consumption and mitigate the costly intervention in network restructuring and reconfiguration. 
The size of heat pump installation is a trade-off between the price of new heat pumps, the performance of new heat pumps, and the investment costs in new lines. Depending on the change in the market price one decision might off-set another in terms of operational costs. Another factor is the change in the consumption volume. If the volume of demand is likely to change in coming decade, a mix of different energy resources might present a sustainable solution to meet the rising demand. However, the condition of network would determine the optimal portfolio which is cost effective and technically feasible. Upgrading the whole network is a demanding  investment both technically and economically which is typically not practical. Therefore this paper sets the objective to find a balance between network interventions to maintain the condition with optimal portfolio of integrated energy resources.

\subsubsection{The challenges of network condition}

The expansion planning of power network can be summarized into two parts: the condition of towers and the condition of lines. The associated power apparatus are part of the tower or line depending on the where the apparatus is installed. A health index as introduced in \cite{manninen2021health} is used for representing the health of towers. In addition to that, a risk factor associated to a region to which the tower is serving is also applied. The risk factor is essentially a weight that the system operator allocates depending on the specific consumer segment. A fragility curve metric is used in literature to determine the health of a transmission tower \cite{raj2021fragility, ma2021component} 

Transmission lines in power networks are supported by towers primarily for long distance power transmission. The longer the transmission line is, the higher the voltage and ground clearance level are. To maintain a safe tension in the transmission lines, a permissible limit of sag (line bending) is adopted. 
When the condition of transmission towers deteriorate, the level of transmission lines between the support towers bends. This in turn lowers the line tension violating the effective ground clearance. Excess sag leads to increase in power losses and may also lead to a potential power failure. In addition, the amount of conductor needed to transfer the required amount of power also increases, as does the cost. Typically the network operator resolves the aforementioned issue by re-instating the line tension through increasing the tower height (transmission level) or by stretching the line from both ends (distribution level).
To avoid this, effective level of sag with vertical ground clearance \cite{goswami2011minimization} must be maintained. Additional transmission or support towers may also be needed to maintain the line levels if the condition of a certain tower has worsened. Effectively identifying the tower condition and maintaining the health would keep the power network safe and ensure an efficient power flow. Various transmission tower conditions through health index are presented in \cite{manninen2021toward}.


\subsection{Objectives and contributions}

The main objective of this paper is to expand the techno-economic optimization tool with additional features presented in \cite{bordin2021multihorizon} to enable the following tasks within the decision making process:
\begin{itemize}
    \item Introduce a novel condition of network through the health index of towers, associated risk and maintenance requirement of lines. 
    \item Experiment and identify how the condition of power network influence the generation and transmission expansion planning of a virtual power plant in an energy system.
    \item Define through sensitivity analysis how heat pump as a local aggregated energy source can balance the operational costs and reduce costly investments in network restructuring and reconfiguration.  
    \item Illustrate through trade-off between the decision variables (network interventions, new capacity installations) and key parameters (condition of network, COP of heat pump, health of towers and associated risk factors, cost of energy resources), the value of change considering overall system gains. 
\end{itemize}

The key contribution and novelties of this paper can be summarized as follows:
\begin{itemize}
    \item Represent a virtual power plant in an integrated electric and heat energy system through dispatchable generation and heat pump units.
    \item Introduce a novel framework representing the conditions of power networks (through health index of tower, risk associated with each tower, and maintenance requirements of lines) in a way suitable for inclusion within decision support systems tools based on mathematical optimization. 
    \item Facilitate a more practical representation of the physical network on which the virtual power plant operates through a mathematical optimization model for resilient expansion planning. 
    \item Introduce risk factors associated to a tower which is based on type of consumer being served by that tower. 
    \item Develop novel mixed integer linear programming features to replace towers and/or lines in a virtual power network as part of the optimal decision.
    \item Reflect real world planning horizon considering lines, towers and heat pump through multi-horizon decision making for network reinforcement and capacity building.
\end{itemize}

In addition to that, extensive sensitivity analysis are conducted as part of the analytical solutions to highlight the trade-offs in decision making. Including the trend of cost, demand growth projections among the parameters; and multiple network conditions from critical to normal depending on the health indices.   



This paper is structured as follows: Section 1 establishes the context and key challenges along with overview of the literature; Section 2 describes the proposed mathematical model and features; Section 3 expands the experiments and discussions on the results; Section 4 concludes with future remarks.

\section{Methodology}

This paper builds on the mathematical optimization models proposed in \cite{bordin2021multihorizon} and \cite{mishra2019rnr} by including key novel features that enhance the decisions making process and expand the possibilities for sensitivity analyses. The novel modelling features refer to:

\begin{itemize}
    \item modelling the health of towers and the possibility to replace nodes;
    \item modelling the heat demand and the possibility to allocate new heat pump capacity.
\end{itemize}

The fundamental structure of the model remains the one proposed in \cite{bordin2021multihorizon}. Therefore the modelling of network restructuring and reconfiguration, as well as lines maintenance costs in a multihorizon perspective, and linearized power flow will not be discussed in this paper. The reader is invited to first familiarize with the modelling foundations proposed in \cite{bordin2021multihorizon} to enhance the understanding of the additional features proposed in this follow up paper. 
The nomenclature proposed in this paper is the same as the one proposed in \cite{bordin2021multihorizon}. Only the additional constraints and modelling features will be discussed in this work, together with the related additional variables and parameters. The constraints and features discussed in this paper works holistically with the constraints and features already thoroughly presented in \cite{bordin2021multihorizon}.

\subsection{Health of towers and nodes replacement decisions}

As outlined in the previous sections, one relevant challenge in power networks is related to the condition of lines and the condition of towers. While the condition of lines has been thoroughly modelled and investigated in \cite{bordin2021multihorizon}, the condition of towers is a feature that is currently completely overlooked in literature, when it comes to energy and power systems mathematical optimization models. 

When studying a power network as a graph with arcs and nodes, we assume that arcs represent power lines, while nodes represent towers. If the state of health of a tower degrades, this has an impact on the performance of the connected lines. If the tower condition is too poor, even a brand new line would not perform at its best, due to the negative impact that the connected tower may have on the line's properties. Therefore, a relevant trade-off arises between the need of replacing lines and the need of replacing towers. In graph theory, the need of replacing towers corresponds to the task of replacing nodes. To the best of the author's knowledge, there are no works in literature that provide modelling formulations suitable to replace nodes within mathematical optimization models, to tackle reliability oriented network restructuring issues of power systems.

The state of health of a tower in a node $i$ is identified by a new parameter $V^{health}_{i}$, while the decision of replacing a tower in a node $i$ is identified by a binary variable $\gamma_{i}$ that is equal to 1 if a pole is replaced on node $i$, or zero otherwise.

The towers replacement is tightly connected with the decisions that involves both existing and new lines. Existing lines can be kept as they are or replaced with new better ones. In addition, new potential lines can be built between towers. This has been modelled in \cite{bordin2021multihorizon}.
However, the performance of both existing lines and new lines is affected by the state of health of towers and by the decisions that involve potential towers' replacement.

The following paragraphs will outline how the towers' state of health and replacement can be modelled as a novel feature within \cite{bordin2021multihorizon}.

\begin{scriptsize}
\begin{equation} \label{eq:pole01E}
\centering
{\gamma_{i} = 1 \Rightarrow p_{i,j,t,a} <= \overline E_{i,j}*V^{health}_{j}*(1- \gamma_{j}) + \overline E_{i,j}* \gamma_{j} 
\qquad \forall (i,j,t,a)|E_{i,j}=1; X_{i,j}=0
}
\end{equation}

\begin{equation} \label{eq:pole03E}
\centering
{\gamma_{i} = 0 \Rightarrow p_{i,j,t,a} <= \overline E_{i,j}* \frac{V^{health}_{i} + V^{health}_{j}}{2} *(1-\gamma_{j}) + BigM * \gamma_{j} 
\qquad \forall (i,j,t,a)|E_{i,j}=1; X_{i,j}=0
}
\end{equation}
\end{scriptsize}

Constraint \ref{eq:pole01E} and \ref{eq:pole03E} work for existing lines. They impose that, if a tower is not replaced, its state of health will penalize the capacity of the line that is connected to the tower. In particular, given a power line represented by an arc $i-j$, if a tower on node $i$ is replaced but a tower on node $j$ is not replaced, then the line capacity is penalised by the state of health of tower $j$ (constraint \ref{eq:pole01E}). 

The same constraint \ref{eq:pole01E} can be rewritten by swapping $i$ and $j$. This way the mirroring condition is imposed: if a tower on node $i$ is not replaced and a tower on node $j$ is replaced, the line capacity is penalised by the state of health of the tower on node $i$.

If a tower on node $i$ is not replaced and a tower on node $j$ is not replaced, the line capacity is penalised by the average value of the health of the two towers (constraint \ref{eq:pole03E}). This latter condition assumes that an average value of the towers' health can be utilized to penalize the line capacity. Of course, more refined representations can be utilized in the model, if industrial real-world expertise suggests different health functions involving more than one tower.

The proposed equations have been modelled using indicator constraints. A similar approach to if-then formulations has been successfully used also in \cite{bordin2019smacs} and \cite{bordin2019including}. For further reading about handling indicator constraints in mixed integer problems see \cite{belotti2016handling}.

\begin{scriptsize}
\begin{equation} \label{eq:pole01R}
\begin{aligned}
\gamma_{i} = 1 \Rightarrow p_{i,j,t,a}<= {} & \overline E_{i,j} * (1 - \sum_{c,a1=1}^{a1=a-Z-1}  k_{i,j,c,a1}) \\
& + (\sum_{c,a1=1}^{a1=a-Z-1} k_{i,j,c,a1} * \overline N_c)*V^{health}_{j}*(1- \gamma_{j}) \\
& + (\sum_{c,a1=1}^{a1=a-Z-1} k_{i,j,c,a1} * \overline N_c)* \gamma_{j} \\
& \forall (i,j,t,a)|E_{i,j}=1; X_{i,j}=1
\end{aligned}
\end{equation}

\begin{equation} \label{eq:pole03R}
\begin{aligned}
\gamma_{i} = 0 \Rightarrow p_{i,j,t,a}<= {} & \overline E_{i,j} * (1 - \sum_{c,a1=1}^{a1=a-Z-1}  k_{i,j,c,a1}) \\
& + (\sum_{c,a1=1}^{a1=a-Z-1} k_{i,j,c,a1} * \overline N_c)*\frac{V^{health}_{i}
+ V^{health}_{j}}{2} *(1-\gamma_{j}) \\
& + (\sum_{c,a1=1}^{a1=a-Z-1} k_{i,j,c,a1} * \overline N_c)* BigM* \gamma_{j} \\
& \forall (i,j,t,a)|E_{i,j}=1; X_{i,j}=1
\end{aligned}
\end{equation}
\end{scriptsize}

Constraint \ref{eq:pole01R} handles the towers replacement when also restructuring decisions are involved. If a tower on node $i$ is replaced and a tower on node $j$ is not replaced, then not only the capacity of existing lines is penalized, but also the capacity of new lines (that may be installed to replace existing obsolete lines) has to be penalized by the state of health of the tower on node $j$. This means that, if an existing line is replaced by a new line ($k_{i,j,c,a} =1$), then the capacity of the existing line is penalized for the years that come before the beginning of the construction time, while the capacity of the new line is penalized for the years that come after the construction time. 

The same constraint \ref{eq:pole01R} can be rewritten by swapping $i$ and $j$. This way the mirroring condition is imposed if a tower on node $i$ is not replaced and a tower on node $j$ is replaced.

Constraint \ref{eq:pole03R} handles the towers replacement when restructuring decisions are involved and when both towers on node $i$ and $j$ are not replaced. Again this condition assumes that an average value of the towers' health can be utilized to penalize the line capacity. The new constraints handle the penalization of existing lines before the construction time and the penalization of new lines aimed at replacing existing lines after the construction time.

The constraints proposed in \ref{eq:pole01R} and \ref{eq:pole03R} are non-linear since they involve the product of two binary decision variables $k_{i,j,c,a}$ and $\gamma_{j}$.
Such constraints have been properly linearized by creating a new binary variable $\mu_{i,j,t,a} = k_{i,j,c,a} * \gamma_{j}$ and by including three inequalities in the form of:

\begin{scriptsize}
\begin{equation} \label{eq:linear01}
\centering
{\mu_{i,j,t,a} \leq k_{i,j,c,a}
}
\end{equation}

 \begin{equation} \label{eq:linear02}
\centering
{\mu_{i,j,t,a} \leq \gamma_{j}
}
\end{equation}
  
 \begin{equation} \label{eq:linear03}
\centering
{\mu_{i,j,t,a} \geq k_{i,j,c,a} + \gamma_{j} -1
}
\end{equation}
\end{scriptsize}

The new binary variable $\mu_{i,j,t,a}$ can then be used inside the non-linear constraints to replace the product of the binary variables $k_{i,j,c,a}$ and $\gamma_{j}$. Of course each constraint has to be linearized individually.

\begin{scriptsize}
\begin{equation} \label{eq:pole01P}
\begin{aligned}
\gamma_{i} = 1 \Rightarrow p_{i,j,t,a} <= {} & (\sum_{c,a1=1}^{a1=a-Z-1} y_{i,j,c,a1} * \overline N_c) *V^{health}_{j}*(1- \gamma_{j}) \\
& + (\sum_{c,a1=1}^{a1=a-Z-1} y_{i,j,c,a1} * \overline N_c) * \gamma_{j} \\
& \forall (i,j,t,a)|N^{pot}_{i,j}=1
\end{aligned}
\end{equation}

\begin{equation} \label{eq:pole03P}
\begin{aligned}
\gamma_{i} = 0 \Rightarrow p_{i,j,t,a} <= {} & (\sum_{c,a1=1}^{a1=a-Z-1} y_{i,j,c,a1} * \overline N_c) *\frac{V^{health}_{i} + V^{health}_{j}}{2}*(1- \gamma_{j}) \\
& + (\sum_{c,a1=1}^{a1=a-Z-1} y_{i,j,c,a1} * \overline N_c) * \gamma_{j}*BigM \\
&  \forall (i,j,t,a)|N^{pot}_{i,j}=1
\end{aligned}
\end{equation}
\end{scriptsize}

Constraints \ref{eq:pole01P} and \ref{eq:pole03P}, \eqref{eq:pole03P_1}, \eqref{eq:pole03P_2} handle the towers replacement when new potential installations are involved. They work similarly to the constraints presented for the restructuring decisions, hence the same considerations made in the previous paragraphs in terms of linearization and indicator constraints apply.

In addition to the constraints proposed above, the decisions in terms of replacing towers affect also the objective function and investment costs. In particular, a new term of cost is created as follows:

\begin{scriptsize}
\begin{equation} \label{eq:pole03P_1}
\centering
{\sum_i V^{CRF} * \gamma_{i}*V^{cost}}
\end{equation}
\end{scriptsize}

Where $\gamma_{i}$ is the binary decision variable for new towers installation, $V^{cost}$ is the installation cost of new towers, and $V^{CRF}$ is the capital recovery factor to actualize investments in new towers. The latter is defined as:

\begin{scriptsize}
\begin{equation} \label{eq:pole03P_2}
\centering
{\frac{r*(1+r)^{V^{life}}}{(1+r)^{V^{life}}-1}}
\end{equation}
\end{scriptsize}

\subsection{Handling heat demand}

Compared to the work presented in \cite{bordin2021multihorizon}, this paper aims at a more detailed representation of the final demand, by considering two separate aggregated demand curves at each node: thermal demand and electrical demand. Regarding the aggregated thermal demand, for residential areas, we refer mainly to low temperature demand for space heating and water heating (namely room heating, snow melting, floor heating, shower heating). For industrial areas we refer mainly to high temperature demand for process heat in industry. As for the aggregated electrical demand we refer mainly to loads such as heating or cooling load through air conditioning, power appliances (i.e. computers, fridge, tv, fan, lighting load etc), electric vehicles charging.
The thermal demand can be met in two main ways: traditional heating systems (namely boilers, district heating, wood/pellets burning) and power-to-heat technologies. The latter refers to the conversion of electrical energy to heat, where the main technology is currently represented by heat pumps. Due to the higher efficiency compared to traditional electric resistive heaters,  flexible heat pumps are an important aspect of flexibility. Therefore their integration can be key to fulfill an increased portion of energy demand without the need for expensive network reinforcement. 
This paper introduces the decision for new heat pump installation to fulfill part of the heating demand. This allows looking at ways in which heat pumps can bring added flexibility to the system and lower the investments in network restructuring and reconfiguration. Looking at future projection of increased energy demand, we are red in particular in understanding which is the portion of heat demand that should be satisfied with a power-to-heat technology (heat pump) in order to minimize the overall costs of network restructuring and reconfiguration.

\begin{scriptsize}
\begin{equation} \label{eq:heat01}
\centering
{f^{in}_{i,t,a} = D_{i,t,a} + d^{heat01}_{i,t,a} + \frac{d^{heat02}_{i,t,a}}{COP} \qquad \forall (i,t,a)}
\end{equation}

\begin{equation} \label{eq:heat02}
\centering
{d^{heat01}_{i,t,a} + d^{heat02}_{i,t,a} = D^{heat}_{i,t,a} \qquad \forall (i,t,a)}
\end{equation}

\begin{equation} \label{eq:heat03}
\centering
{d^{heat02}_{i,t,a} <= \sum_{a1=1}^{a1 = a} h_{i,a1} \qquad \forall (i,t,a)}
\end{equation}
\end{scriptsize}

Constraint \ref{eq:heat01} defines the power balance in each node. The left side of the constraint summarizes the power flow into the node through the term $f^{in}_{i,t,a}$. This term is split and thoroughly described in \cite{bordin2021multihorizon} and the left side of this constraints remains the same as the one proposed in \cite{bordin2021multihorizon}. The right side of the constraint is different, since the demand curves are now split into two: power demand $D_{i,t,a}$ and heat demand. The latter is further split into the portion of heat demand that can be met by traditional hating systems $d^{heat01}_{i,t,a}$ and the portion of heat demand that can be met through a heat pump $d^{heat02}_{i,t,a}$. The latter is divided by the heat pump coefficient of performance $COP$ which can be changed to allow sensitivity analyses during computational experiments. Both $d^{heat01}_{i,t,a}$ and $d^{heat02}_{i,t,a}$ are decision variables, so it is an optimal decision of the model how much heat demand is worth satisfying with existing traditional heating systems and how much heat demand is worth satisfying with new aggregated heat pump installation in that particular node.

Constraint \ref{eq:heat02} defines the total heat demand curve as the summation of the heat demand that is satisfied using existing traditional heating systems and the heat demand that is satisfied by installing new heat pump capacity

Constraint \ref{eq:heat03} keeps track of the heat pump aggregated capacity installation in each node and for each strategic year. For each strategic year, the heat pump capacity in a node is given by the heat pump installed in the previous years plus the capacity installed in the current year. The portion of heat demand satisfied by heat pumps should not exceed the available installed capacity.

Finally, an additional term in the objective function must be added to include the investment costs in new heat pump capacity installation. Again, the capital recovery factor is used and the related formula is given in the following equation \ref{eq:heat04} where $H^{CRF}$ is the capital recovery factor of a heat pump, $h_{i,a}$ is the decision variable related to the installed heat pump capacity, $H^{cost}$ is the unitary cost of a heat pump.

\begin{scriptsize}
\begin{equation} \label{eq:heat04}
\centering
{ \sum_{i,a} H^{CRF}*h_{i,a}*H^{cost}}
\end{equation}
\end{scriptsize}

The capital recovery factor of a heat pump depends on the r rate $r$ and on the heat pump average life $H^{life}$ and it is calculated as in \ref{eq:heat05}

\begin{scriptsize}
\begin{equation} \label{eq:heat05}
\centering
{ \frac{r*(1+r)^{H^{life}}}{(1+r)^{H^{life}}-1} }
\end{equation}
\end{scriptsize}

\subsection{Introducing a risk factor}

 A risk factor is introduced in the mathematical optimization model, to prioritize investments in areas that are considered as more critical for the power system companies. The fundamental idea behind a risk factor is that areas with a high risk have a higher priority for the power system company, and therefore tighter constraints in terms of meeting the existing and new forecast demand. While areas with a lower risk have a lower priority for the power system company, which would allow relaxing the constraints in terms of meeting the existing and new forecast demand, by allowing for some demand not met. This will of course affect the decisions in terms of lines restructuring and reconfiguration as well as poles replacement, since low risk areas can still be connected to obsolete lines or towers allowing for a portion of demand not met. The risk factor is changing over the years, since the priority of a certain node might increase or decrease in the future. For instance, if a new district is planned in a node that is currently not highly populated, the priority of that particular node may increase in the forthcoming years once the district will be fully developed. If an area changes its features or the population is gradually moving out, the priority for that particular node may decrease in the forthcoming years. 
 
The risk factor is included in the model by using a risk parameter $V^{risk}_{i,a}$ in the power flow equation as shown in \ref{eq:risk}. It appears in the model as a percentage applied to the total demand in the node. 

\begin{scriptsize}
\begin{equation} \label{eq:risk}
\centering
{f^{in}_{i,t,a} = (D_{i,t,a} + d^{heat01}_{i,t,a} + \frac{d^{heat02}_{i,t,a}}{COP})*V^{risk}_{i,a} \qquad \forall (i,t,a)}
\end{equation}
\end{scriptsize}

However, it should be highlighted that the risk factor is much more than just a single number, and that there is a whole set of qualitative and quantitative analyses that have to be performed to properly define it. 
In particular, the following definition of risk factor applies:

\begin{scriptsize}
\begin{equation} \label{eq:riskdef}
\centering
{V^{risk}_{i,a} = f [(F^{ex}_{1,c}, F^{ex}_{2,c}, ... F^{ex}_{n,c}) ; (F^{in}_{1,c}, F^{in}_{2,c}, ... F^{in}_{n,c});  (F^{ex}_{1,x}, F^{ex}_{2,x}, ... F^{ex}_{n,x}) ; (F^{in}_{1,x}, F^{in}_{2,x}, ... F^{in}_{n,x})]} \end{equation}
\end{scriptsize}

Where the risk factor is a function of different factors as follows:
\begin{itemize}
\item $F^{ex}_{1,c}, F^{ex}_{2,c}, ... F^{ex}_{n,c}$ is a function of external controllable factors; 
\item $F^{in}_{1,c}, F^{in}_{2,c}, ... F^{in}_{n,c}$ is a function of internal controllable factors; 
\item $F^{ex}_{1,x}, F^{ex}_{2,x}, ... F^{ex}_{n,x}$ is a function of external non controllable factors;
\item $F^{in}_{1,x}, F^{in}_{2,x}, ... F^{in}_{n,x}$ is a function of internal non controllable factors.
\end{itemize}

The above definition explains that the risk is a function of different factors $F$ that affect the overall classification of a certain area. Examples of such factors are for instance the terrain, the number of consumers, the type of consumers, the seasonality of consumers availability (i.e. if consumers are present all over the year or just in winter or summer), the future projections for regional development (i.e. future demand projections and future projections in terms of new infrastructures development), type of nature that characterizes a node (i.e. if a node is surrounded by a forest it may be difficult to reach), distance from water, weather patterns (i.e. length and harshness of certain seasons such as winter or summer).
The factors affecting the conditions of power networks can be broadly classified into two types: controllable and uncontrollable. While both can be monitored, the controllable factors can be altered while the uncontrollable factor can be slowed down. For example, a sag formation on the overhead transmission line can be classified as a controllable factor as it can be altered through tightening the lines and increasing the tower height. A material deterioration of a transmission line, on the other hand, cannot be controlled as it is a natural degrading process.
Factors can be also classified as internal or external based on the reason of an issue. For example an over voltage issue can be an external fault that might be caused by high share of renewable. It can be classified as an internal fault if adequate voltage support are not planned by the responsible system operator in the region. Note that the internal and external classification is tightly linked to the cause of a factor (namely, why it happened) while the controllable and uncontrollable classification is tightly linked to the effect of it (namely, what can be done about it).

Each factor in \ref{eq:riskdef} can be then multiplied by a weight $W$ that defines how much that particular factor is important for that particular area and incorporates also information about the probability of occurrence of that factor. So that equation \ref{eq:riskdef} is further expanded into \ref{eq:riskdef2} as follows:

\begin{scriptsize}
\begin{equation} \label{eq:riskdef2}
\begin{aligned}
V^{risk}_{i,a} = {} & f [(F^{ex}_{1,c}*W^{ex}_{1,c}, F^{ex}_{2,c}*W^{ex}_{2,c}, ... F^{ex}_{n,c}*W^{ex}_{n,c}) ;
(F^{in}_{1,c}*W^{in}_{1,c}, F^{in}_{2,c}*^{in}_{2,c}, ... F^{in}_{n,c}*W^{in}_{n,c}); \\
& (F^{ex}_{1,x}*W^{ex}_{1,x}, F^{ex}_{2,x}*W^{ex}_{2,x}, ... F^{ex}_{n,x}*W^{ex}_{n,x}) ;
(F^{in}_{1,x}*W^{in}_{1,x}, F^{in}_{2,x}*W^{in}_{2,x}, ... F^{in}_{n,x}*W^{in}_{n,x})] 
\end{aligned}
\end{equation}
\end{scriptsize}

This is mainly done at a qualitative level by industrial experts. 



\section{Experiments and Findings}

Computational experiments are performed using the proposed model to investigate how the network condition impacts the decision to expand the network capacity with an integrated energy system. Thereafter sensitivity analyses are conducted to illustrate the trade-offs between potential investments in new generation units, heat pump units, network restructuring with lines replacements, and reconfiguration decisions with the dismantling of lines, replacement of towers. 

The VPP and the assets are reflected through an aggregated network. For this purpose, a modified IEEE-9 bus system is used as the base VPP network to run the computational experiments - see fig. \ref{fig:net}. Indeed this network is constructed in a way that it has both non-dispatchable and dispatchable generation units as typically observed in the portfolio of a VPP. 
Realistic data of production and consumption received from the utility industries are used to run the tests. Therefore the construct and data-set closely resemble the real-world setting. The data-set comes with non-disclosure restrictions due to the privacy, commercial interests, ethical and security concerns, which makes it challenging for research publications. 

The objective of this paper is to present a novel methodology to improve the decision-making process for resilient network expansion planning problems while considering the network condition in an integrated energy system. Even though real-world case studies are known to be ``practical", more often than not each problem is unique. Therefore, this paper adopts a validation by construct approach through various sensitivity analyses providing a wider range of examples. For all practical purposes, one real-world case study is also presented in the paper. Beyond this, a validation by construct approach further enables reproduction of results to \cite{flake2021strengthening}. Nevertheless, the authors acknowledge the scope for a comparative analysis based on real network data-set which will be a separate full-length paper in future studies.  

The prototype of the proposed mathematical optimization model is developed using AIMMS modeling platform \cite{bisschop2006aimms}. AIMMS provides a versatile platform to rapidly build and test mathematical optimization models. The experiments presented in the paper are conducted using a personal computer with a 3Ghz processor and 32GB RAM. 

\subsection{Structure of the experiments}

\begin{figure}[H]
    \centering
    \includegraphics[scale = 0.65]{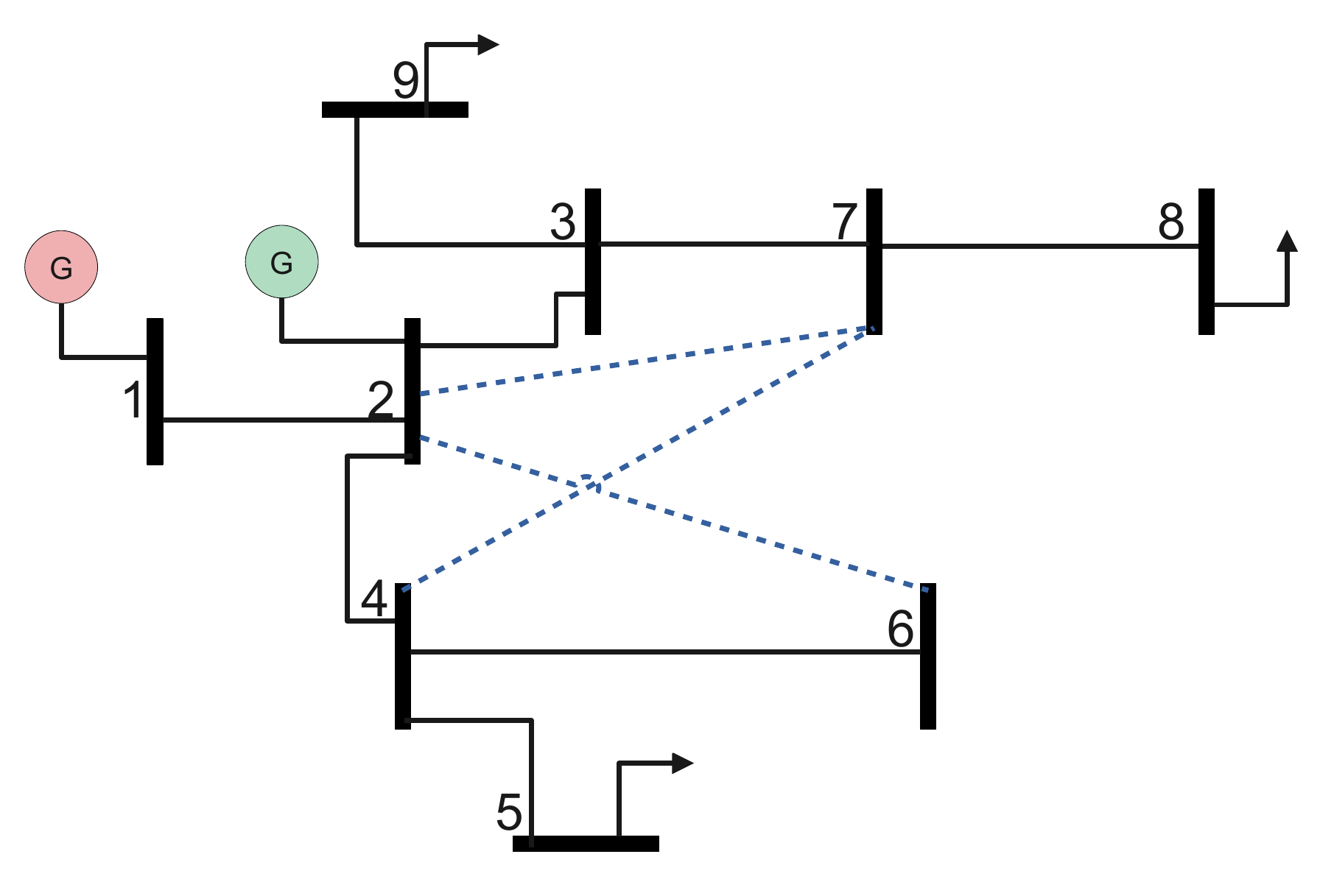}
    \caption{Power network configuration used for computational experiments}
    \label{fig:net}
\end{figure}

The network configuration presented in fig. \ref{fig:net} consists of 9 nodes and 10 arcs. The network is represented at an aggregated level- each node is representative of a group of buildings, each path formed by one or more arcs represents a corridor, a group of arcs and nodes form a zone. The network has both conventional and renewable generation units where a typical off-shore wind turbine production curve is used as the basis for renewable production. The energy consumption in electric and heating demand are presented separately. 
The planning horizon is assumed to be a decade, given that network reconfiguration decisions would typically fall within this period. In addition significant technological changes affecting the unitary cost and consumption trends and patterns could also be observed within this period. The construction time for lines and towers are fixed to three years period, which is a maximum period for construction process under normal conditions. All data are scaled to inherit the pattern and trends while preserving the privacy and commercial interests. 

A conventional generator of sufficient capacity to meet the total load of the grid is in node 1. A variable renewable generation unit, offshore wind turbines, is present in node 2 along with a battery bank. Linearly increasing electric and heat demand are located in node 5, 8, and 9. The total energy demand at node 8 would surpass the line capacities on year-6. The aforementioned assumptions are detailed in a previous paper from the authors in \cite{bordin2021multihorizon}. To reflect the real world choices, the model was given two options, C1 and C2, to choose from when it comes to lines replacement. One of them, C1, is higher capacity and more expensive than the other. 

The following subsections detail the sensitivity analyses for each feature of the model. Every figure includes a set of tables illustrating the decision variables in focus. The decision variables include: reconfiguration through new line installation, restructuring through replacement of existing lines and transmission towers, total investment and operational costs, new capacity building through renewable generation unit or heat pump installations. 
The analyses are meant to understand the effect of various components on the decision variables. The findings focus on the ratio between the level of impact with a range of variations in a component. Thereby the experiments and findings together generalize and widen the understanding on the implications of technology, demand and cost.

\subsubsection{Map of components and challenges}

Small scale non-dispatchable generation assets such as wind turbines and solar panels can be treated as behind the meter solutions. A portion of the electric demand is met locally when these assets are active, while the larger share of demand is sourced from the power grid. Micro injections from small to medium scale non-dispatchable generation to the power grid is considered trivial in terms of volume of power supply. Medium to large scale generation assets such as the off-shore wind turbines have a stable generation curve and smaller storage requirement. To identify the impact of investment in heat pump installations the experiments are broadly classified into with and without existing non-dispatchable renewable power generation assets. A series of experiments are conducted for each category: COP test, heat pump cost with line cost, COP with heat pump cost, increasing and decreasing trends of heating demand. Every experiment begins with a base case accompanied with sensitivity analyses to present the changes.

In the following tables we refer to potential new installation as restructuring tasks.


\subsection{ Network interventions with impact of heat pump}

\begin{figure}[H]
\centering
\subfloat[With non-dispatchable power generation units]{%
  \includegraphics[clip,width=0.8\columnwidth]{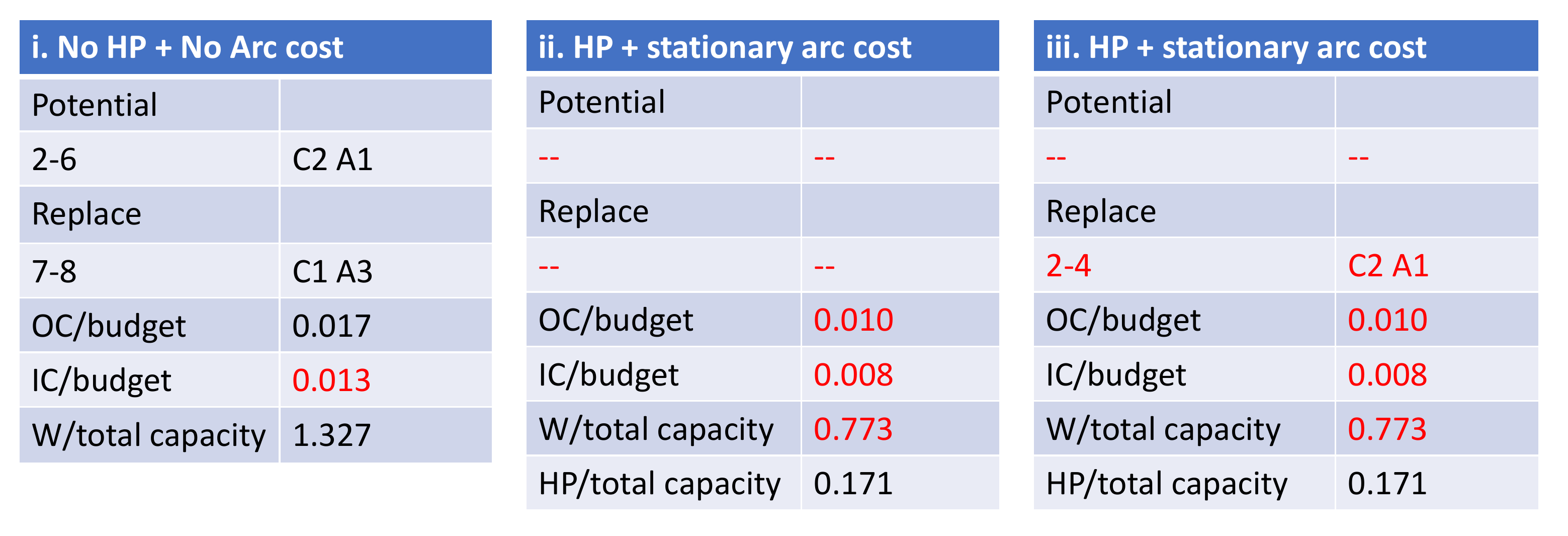}%
    \label{fig:w1}
}

\subfloat[Without non-dispatchable power generation units]{%
  \includegraphics[clip,width=0.8\columnwidth]{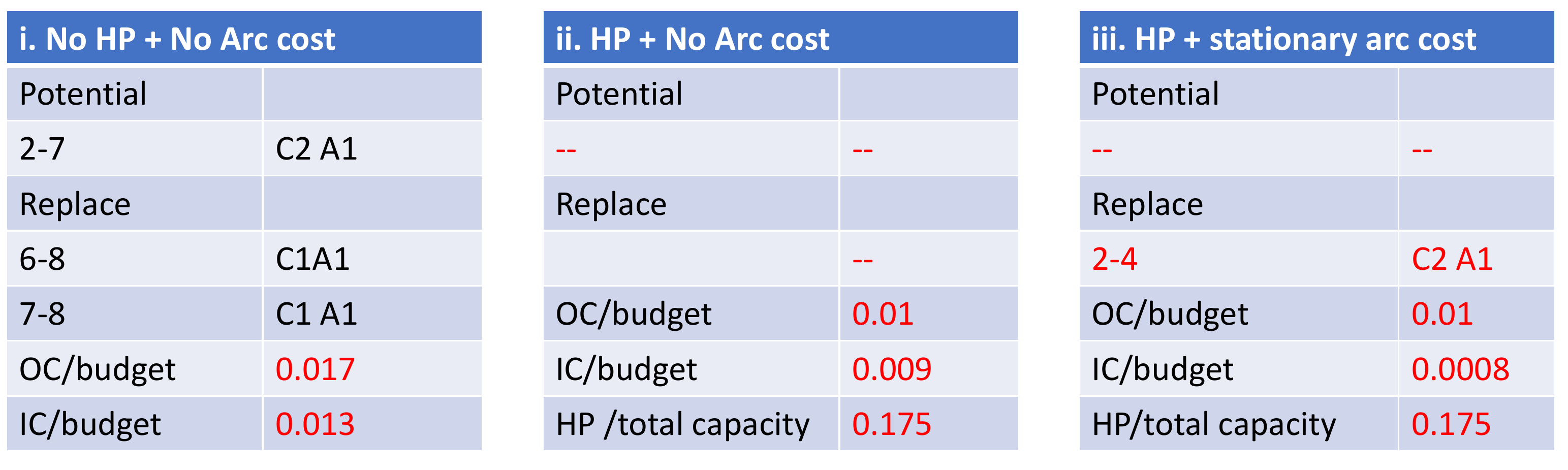}%
    \label{fig:nw1}
}
\caption{Restructuring decisions when heat pump is available}
\label{fig:s1}

\end{figure}

Fig. \ref{fig:w1} presents the model decisions with and without non-dispatchable generation unit. Every table lists the decision variables associated to the case. The condition of the test are explained in the table titles. 
The heating demand at node 8 is projected to surpass the existing line capacities in arcs 7-8 and 6-8 in the sixth year. This creates a congestion in the network, as outlined in the previous section. Therefore reconfiguration and restructuring decisions are observed. 

The underlying assumptions for the set of experiments presented in fig. \ref{fig:w1} are: no maintenance cost of lines and no heat pumps available.

Without availability of heat pump, both reconfiguration (2-6) and restructuring (7-8) decisions are taken to meet the demand. When a heat pump option is available among the potential optimal decisions, since the performance of heat pump is better than traditional electricity to heat conversion, heat pumps are installed in place of any network interventions. Indeed, there is a 39\% decrease in the operational cost and 35\% increase in the investment cost. When the cost to maintain existing lines 2-4 is taken into consideration, then there is a change in the investment scheduling as in- line 2-4 is restructured. Comparing table a.iii with table a.i of fig. \ref{fig:w1} it is possible to note that the new cable replacement in table a.iii is accompanies by no potential installations that we see in table a.i. This is because the possibility to install heat pumps is preferred compared to building new line capacities.


Fig. \ref{fig:nw1} shows the decisions when additional non-dispatchable generation units are not available. Without an option to invest in heat pump units (see table b.i) the optimal decisions are- reconfiguration (2-7) and restructuring (6-8, 7-8). No network interventions are observed when heat pump is available (see table b.ii). Comparing both the previous instances there is a 39\% increase in operational costs and 33.8\% decrease in investment costs with investments in heat pump units. The reason is that the heat pump availability helps avoiding costly investments in network interventions. With inclusion of stationary maintenance cost in arc 2-4 a network intervention decision is taken in form of restructuring (2-4). It can be observed that with investment in heat pump as an option, the number and mode of network interventions significantly reduces even when maintenance cost of lines comes to      the picture. The investment cost increases while the operational cost decreases.

\subsubsection{COP test}

\begin{figure}[H]
\centering
  \begin{subfigure}{0.8\textwidth}
    \includegraphics[width=\textwidth]{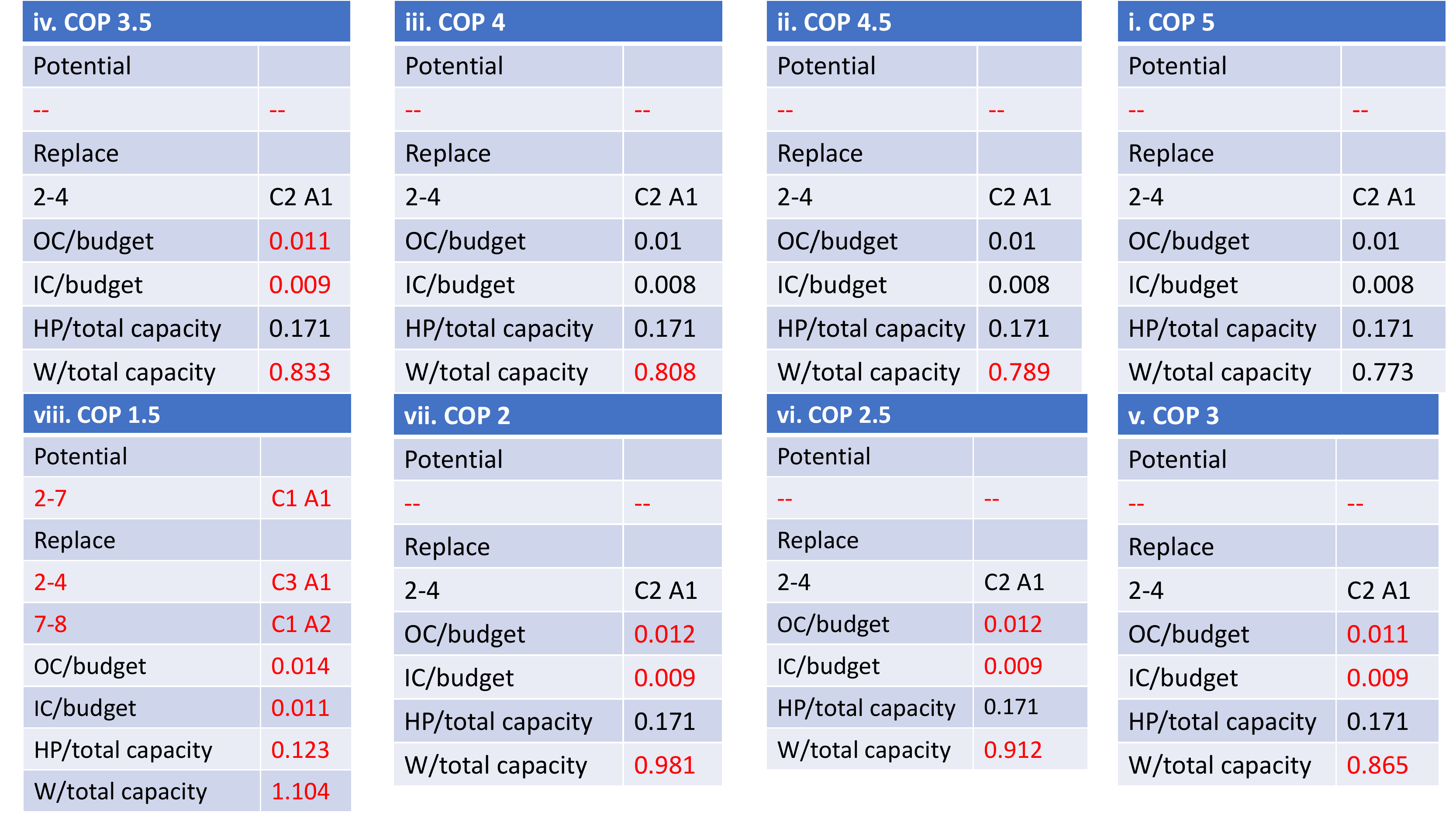}
    \caption{With non-dispatchable power generation units}
    \label{fig:w2}
  \end{subfigure}
  \begin{subfigure}{0.8\textwidth}
    \includegraphics[width=\textwidth]{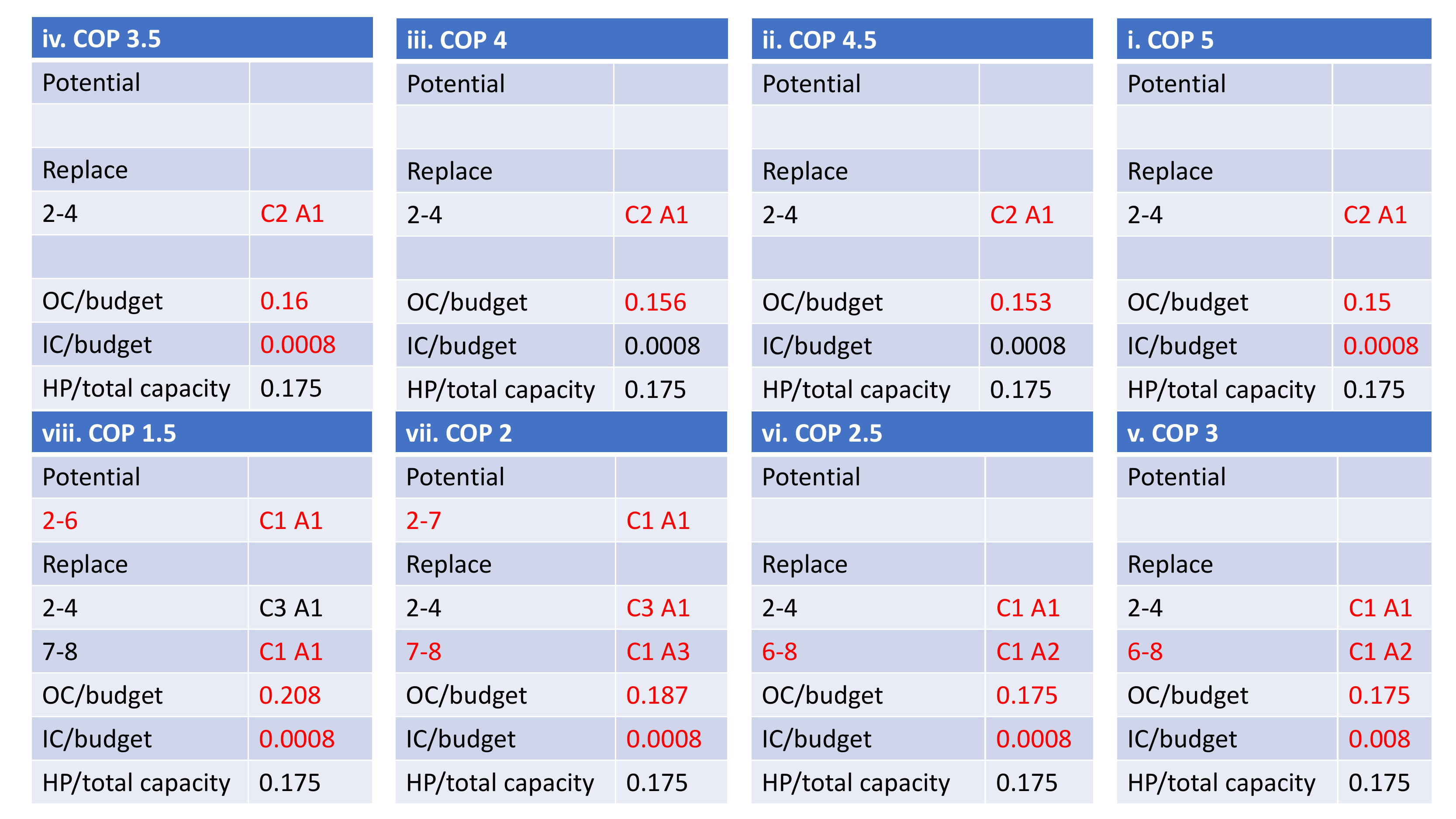}
    \caption{Without non-dispatchable power generation units}
    \label{fig:nw2}
  \end{subfigure}
    \caption{Restructuring decisions with various coefficient of performance of heat pump}
    \label{fig:s2}
\end{figure}

Coefficient of performance of heat pump is an efficiency metric ranging from 5 to 1.5 depending on the technology and other environmental factors\cite{song2021experimental, bae2021comparison}. In \cite{ostergaard2021variable} the authors highlight that COP of 3.5 becomes competitive in spot electricity market prices. 
Fig. \ref{fig:s2} shows the sensitivity analysis of coefficient of performance with and without non-dispatchable generation unit. By gradually decreasing the COP by 10\%, the decision changes in terms of network interventions and costs are presented. 


Comparing the decision changes in both cases, it can be observed that the case without non-dispatchable generation units has more network interventions with decrease in the COP. From COP 2.5 to 2, for instance, results in additional network interventions- reconfiguration (2-7) and restructuring (6-8). Similar network interventions, in case of existing non-dispatchable generation units, are observed from COP 2 to 1.5. This leads to the observation that non-dispatchable generation units introduces additional decision flexibility.

\begin{figure}[H]
    \centering
    \includegraphics[width=0.8\textwidth]{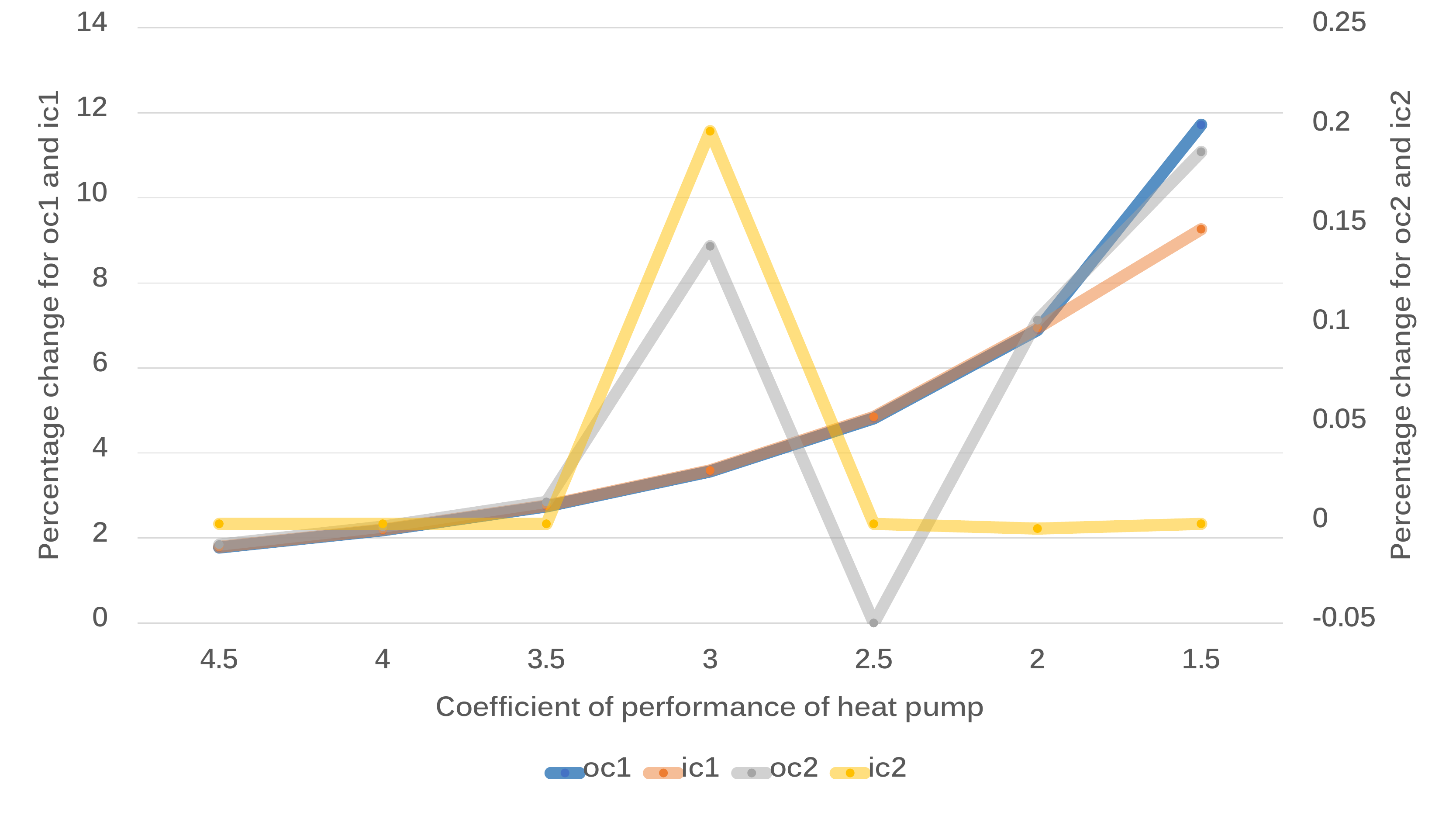}
    \caption{Trend of investment (ic) and operational (oc) costs with descending order of COP of heat pump. With non-dispatchable generation units (ic1, oc1), without non-dispatchable generation unts(ic2, oc2).}
    \label{fig:coptest}
\end{figure}

While the cumulative sum of heat pumps installed over the ten years time horizon remains the same in most of the cases, it was observed that the rate of capacity building over the years are significantly different so as the network interventions. The changes in built heat pump capacities are highlighted in red color. When the COP is assumed as low as 1.5 it can be observed that this is the only case when the cumulative capacity of heat pump installation is lower than the other cases.
Fig. \ref{fig:coptest} details the cost curves for both cases matching the change in cost between two consecutive COP variations. 
Comparing the investment cost curves ic1 and ic2, while ic1 is linearly increasing ic2 is relatively stable with a break-through at COP 3. This means that at COP 3 and beyond the HP becomes an optimal option for investment compared to network interventions. Indeed after COP 3 the change in investment with consecutive increment in COP values remains the same. 
The same trend can be also observed with operational costs (oc1 and oc2). This change is due to the change in efficiency of the heat pump while no flexibility is added from the non-dispatchable generation units. 

Technological innovations often fall back in being cost effective. This in-turn creates a lag in wider adoption of technologies.

\begin{figure}[H]
    \centering
    \includegraphics[width = \textwidth]{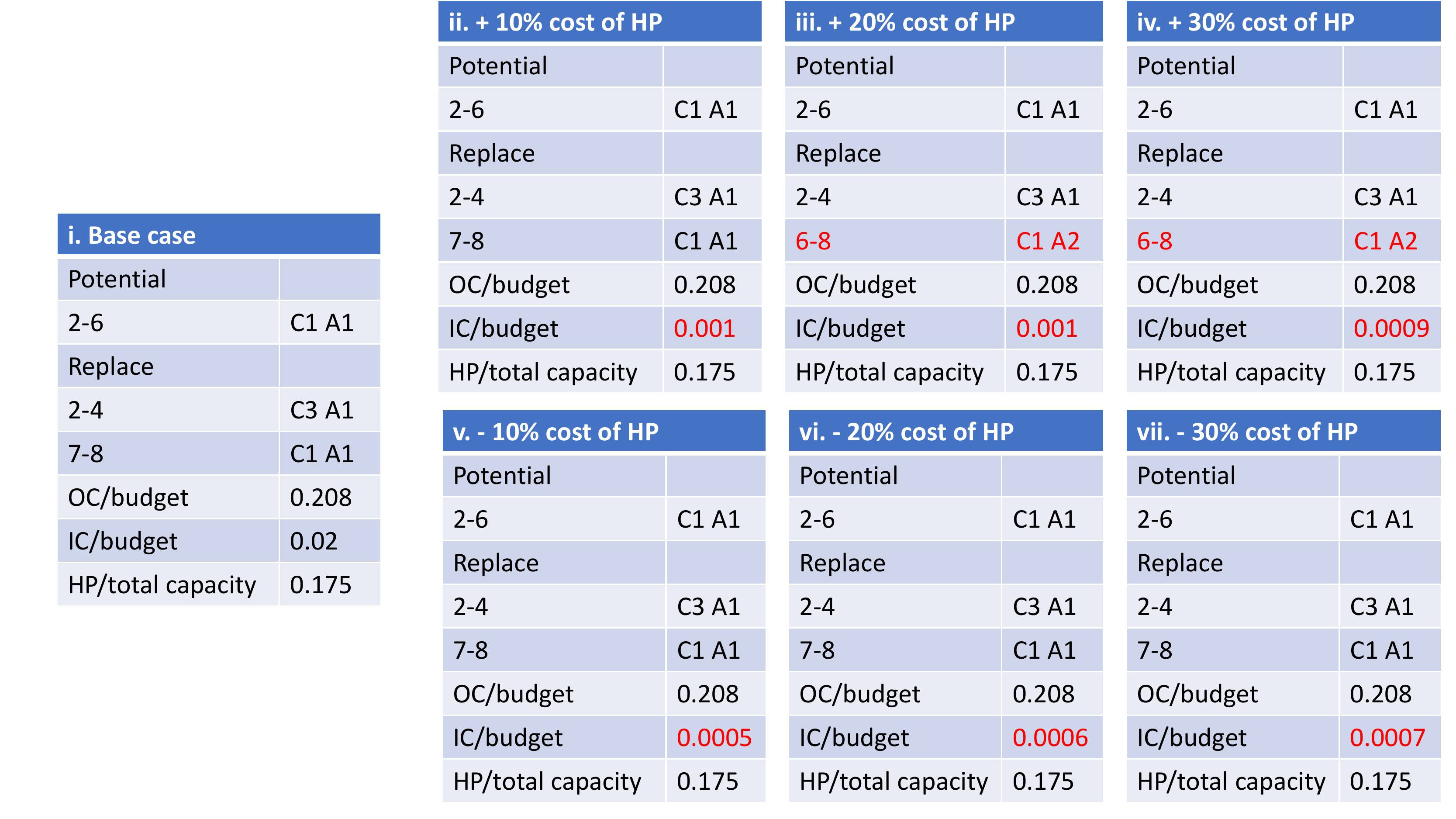}
    \caption{Analysing the ratio between the cost of heat pump and COP. Network interventions without non-dispatchable generation unit with heat pump COP 1.5}
    \label{fig:cop1.5}
\end{figure}

Fig. \ref{fig:cop1.5} illustrates how the ratio between the cost of heat pump and COP influences the network interventions. The reconfiguration decisions are prominent when the cost of heat pump starts to increase in comparison to the base case (1:1). While the total capacity of non-dispatchable generators and heat pumps installed over the time horizon remains the same, it was observed that the rate at which capacity is built throughout the 10 years horizon varies significantly. When the cost of heat pump starts to increase then it is more cost effective to perform network reconfiguration in place of additional heat pump units to meet the demand. While this trend continues, more restructuring decisions are observed in place of additional capacity installations. 

Additional sensitivity analysis have been conducted keeping the COP value constant at 2.5 while the cost of heat pump is gradually decreasing. Particularly when non-dispatchable generation units are an option, the decrease in cost of heat pump does not alter the network intervention decisions as presented before. Indeed there are changes in investment and operational costs which are trivial as change in unitary cost would reflect in overall costs. This is the reason why the results for this particular test are not presented in this paper. 

\subsubsection{HP cost vs New line cost}

\begin{figure}[H]
    \centering
    \includegraphics[width=\textwidth]{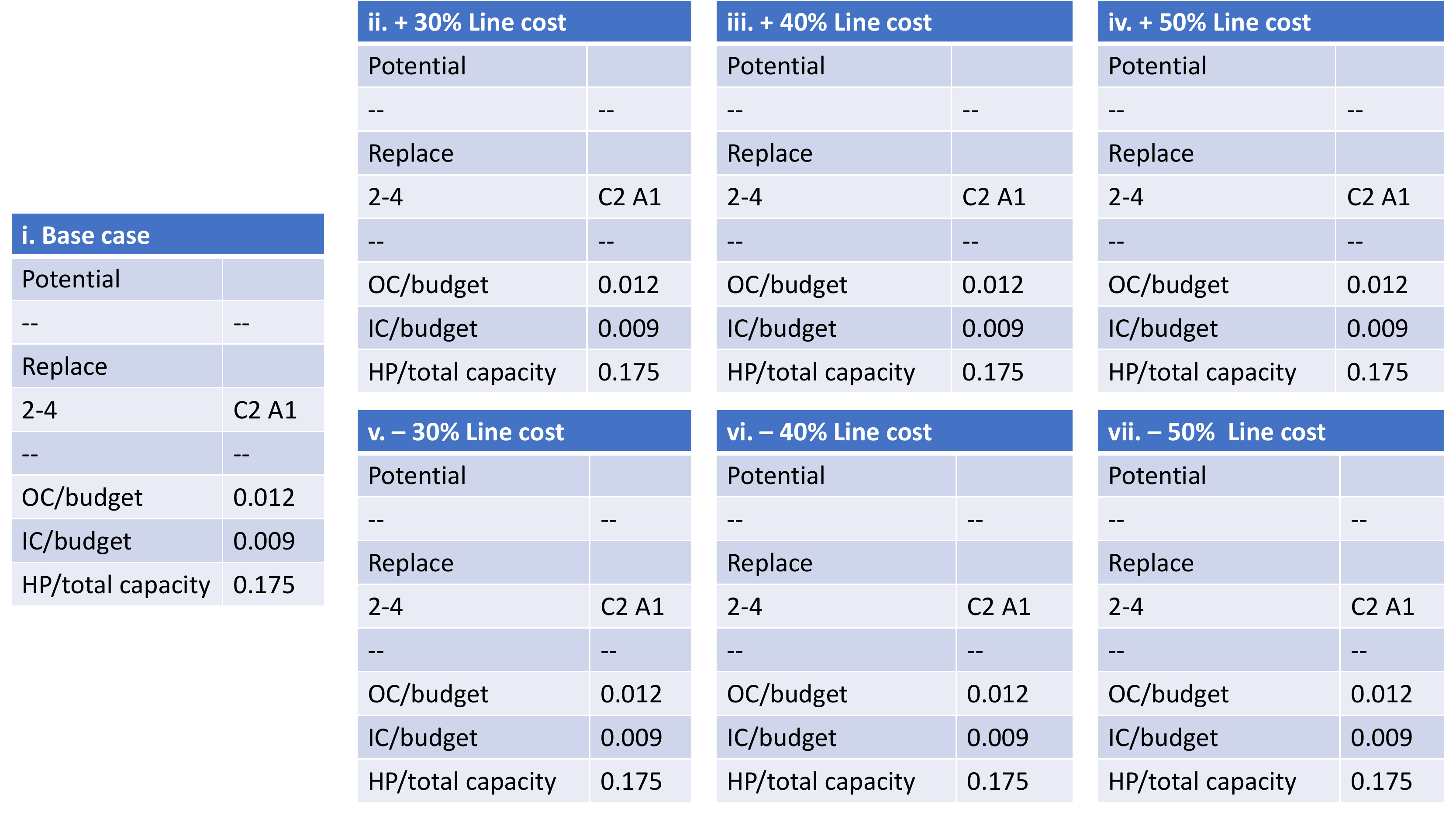}
    \caption{Network interventions comparing cost of heat heat pump and line}
    \label{fig:s3}
\end{figure}

Fig. \ref{fig:s3} shows how the decisions change when the cost of new lines (C1 and C2) changes in comparison with the cost of heat pump. The line types vary based on the cost and capacity, in particular C1 costs more than C2. Fixing the cost of the heat pump, a sensitivity analysis is conducted by increasing and decreasing the cost of the lines. Note that that the COP is set to 2 for this particular case study.

As shown in Fig. \ref{fig:s3} the change in the cable costs affect the investment and operational costs accordingly. While there are no network interventions, the investment and operational costs match the proportional change in the cable costs can be noted. However it was also observed that there is a change in the pattern of installation of heat pump capacities over the time horizon in each case. Expanding on that, when the cost increases/decreases then the model chooses to wait/build capacities accordingly.

\subsubsection{COP vs heat demand}

\begin{figure}[H]
    \centering
    \includegraphics[width=\textwidth]{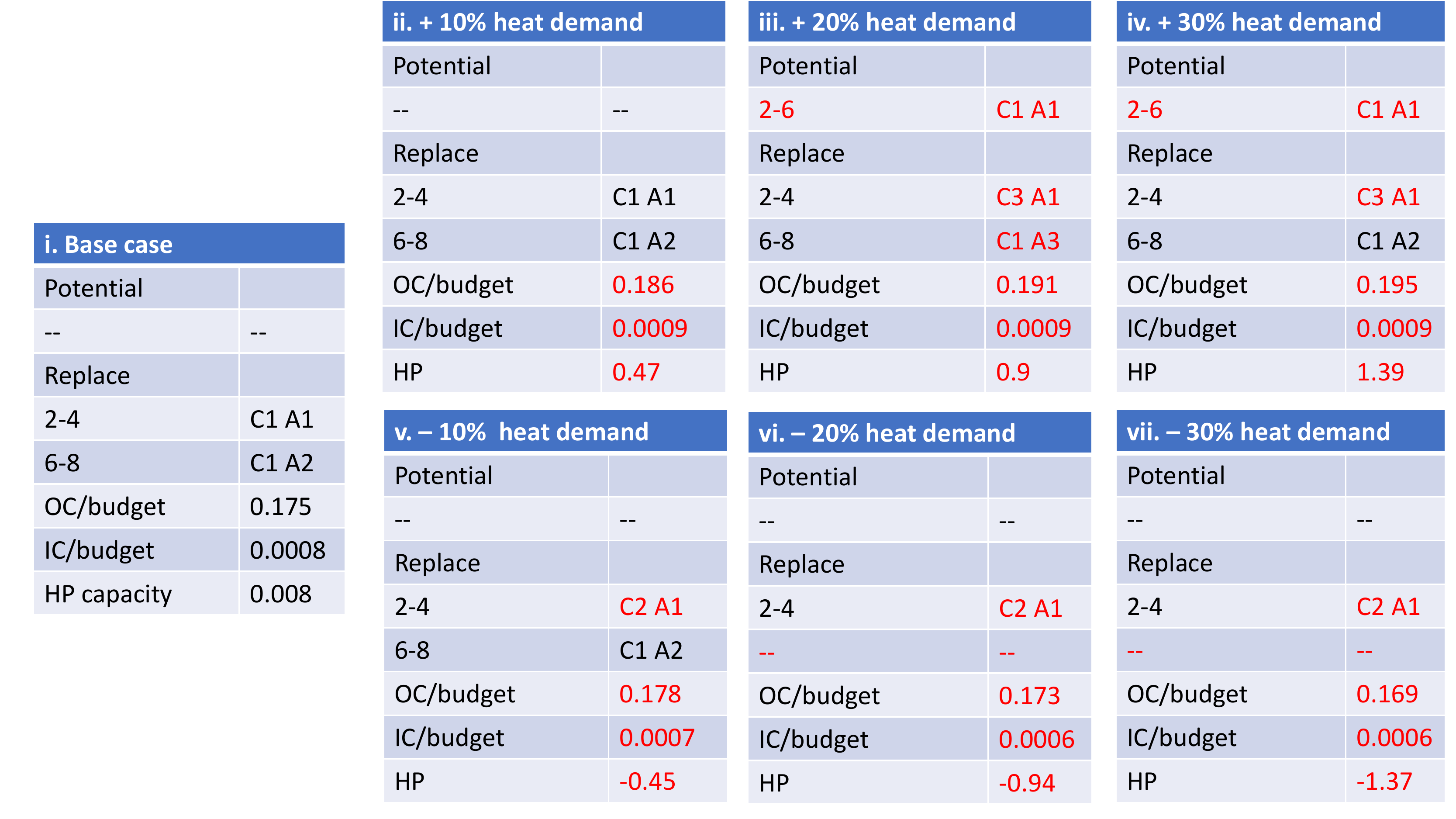}
    \caption{Decisions when comparing COP and change in heat consumption}
    \label{fig:s4}
\end{figure}

Fig. \ref{fig:s4} displays how the decisions change when the heat energy consumption changes. When the demand is increased at the rate of 10\% then new potential line capacities are built to meet the additional demand. 
The total value of heat pump capacity installed for each case is normalized using the the formula $value$ = (data - mean of data)/ standard deviation of the data.

Alongside network restructuring takes place to maintain the power flow. While the heat demand decreases, network restructuring takes place and no new lines capacities are built. 
As opposed to previous cases, when the heat demand changes the total capacity built over the planning horizon changes. Similar to previous cases, the change in the pattern of heat pump capacity building over the years continues. It is evident that a part of the demand is met by the newly built heat pump units and new line capacities to manage the power flow. When the heat demand increases by 20\%, significant network interventions in terms of installing potential lines along with network restructuring are observed. The trend of proportional rise of heat pump capacity with the rise in demand can be clearly observed from the experiment. 

The network interventions are also matched with new heat pump installations to generate the additional power that is required to meet the demand. The operational and investment costs are also increasing over the time horizon. Indeed, installing additional heat pumps is an optimal decision to meet the change in heating demand in comparison to additional extensive and costly network interventions.

\subsubsection{Implication of condition of the tower}

A health index (HI) is utilized to indicate the condition of the transmission tower \cite{manninen2021health}. Lower HI indicates a lower operational capacity. In the proposed model, it translates to a lower transmission capacity. The representative value of HI ranges from 0 to 1, lower the value means lower the associated capacity of transmission lines. A typical transmission tower facilitates multiple overhead lines. Depending the number of lines it facilitates the associated risk changes. A tower acting as a junction to facilitate 3 connections for instance is a high risk tower. Whereas a tower with one connection is low risk since it is serving one consumer. Consumer type (industrial, residential, commercial, office building, emergency care, etc) might also be be used in the risk classification function. In this work, number of connections is used as a risk metric.    

\begin{figure}[!htbp]
  \begin{subfigure}{\textwidth}
    \includegraphics[width=\textwidth]{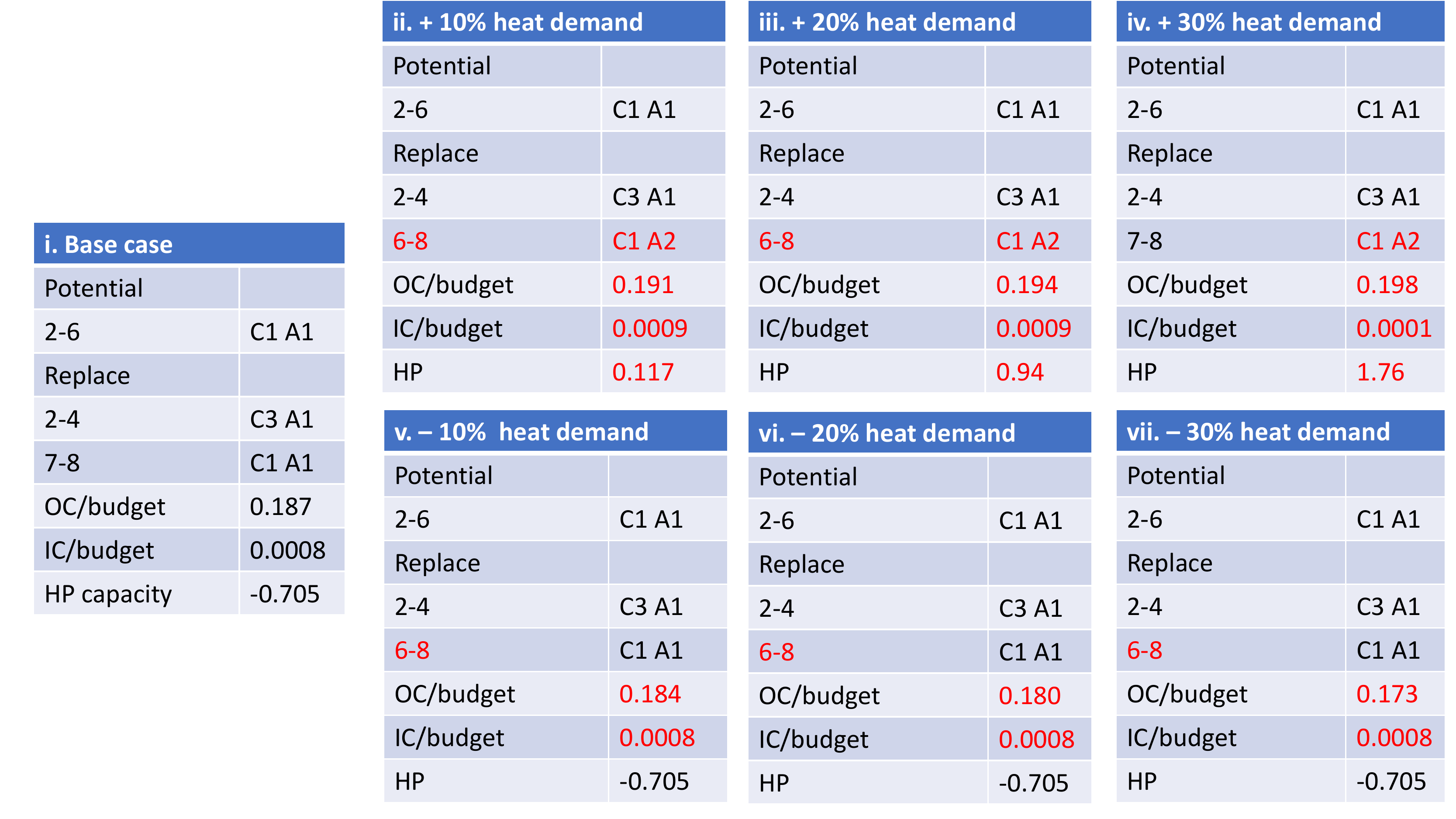}
    \caption{Decisions when heat energy consumption is linearly rising at node 8 and health of Tower 2 is half}
    \label{fig:s5c1}
  \end{subfigure}
  \begin{subfigure}{\textwidth}
    \includegraphics[width=\textwidth]{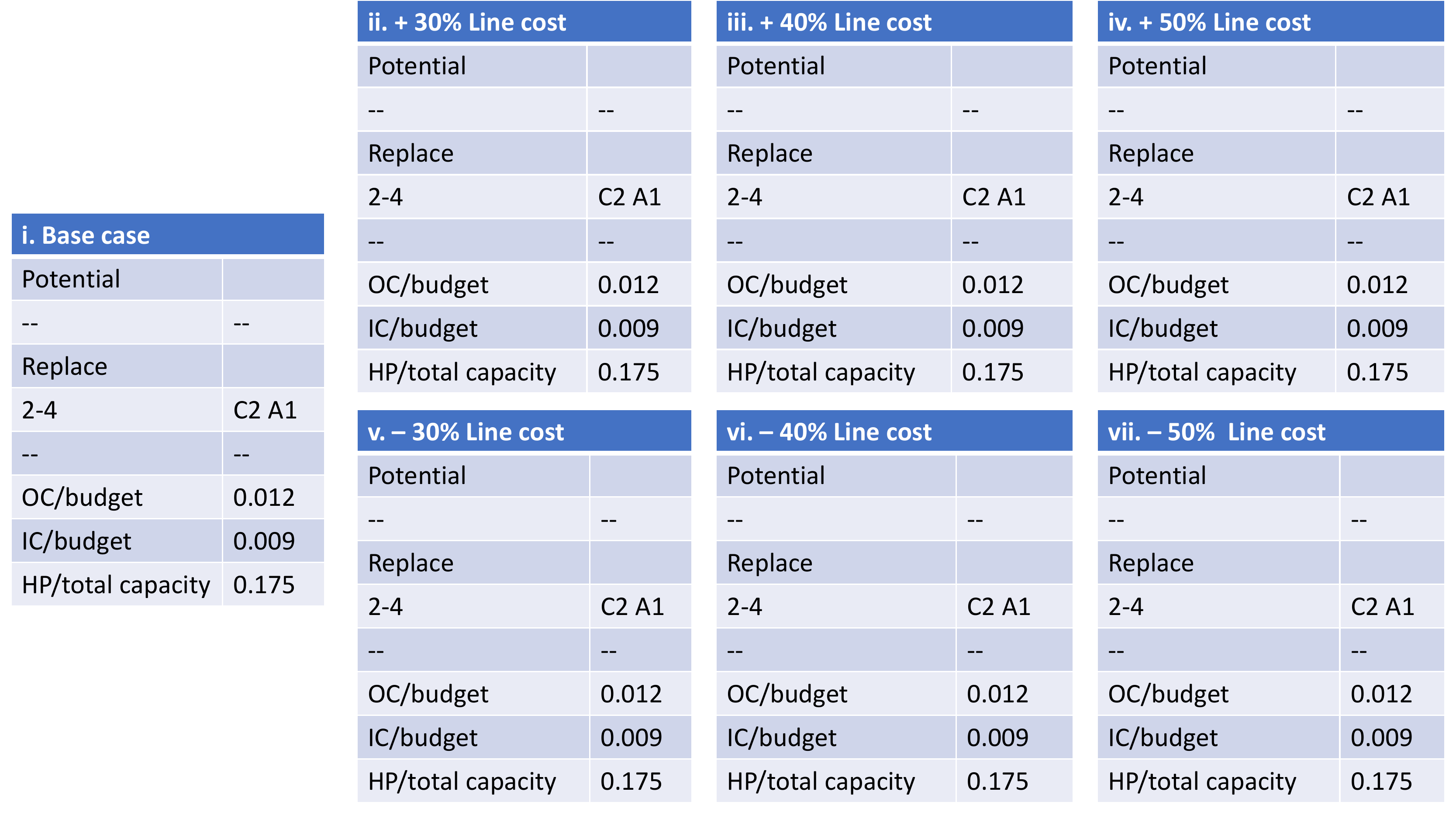}
    \caption{Decisions when the tower 2 condition varies}
    \label{fig:s5c2}
  \end{subfigure}
    \caption{Network interventions and decisions under various (network condition) scenarios of tower 2, 4, and 8}
    \label{fig:s5-1}
\end{figure}
\begin{figure}[H]
  \begin{subfigure}{\textwidth}
    \includegraphics[width=\textwidth]{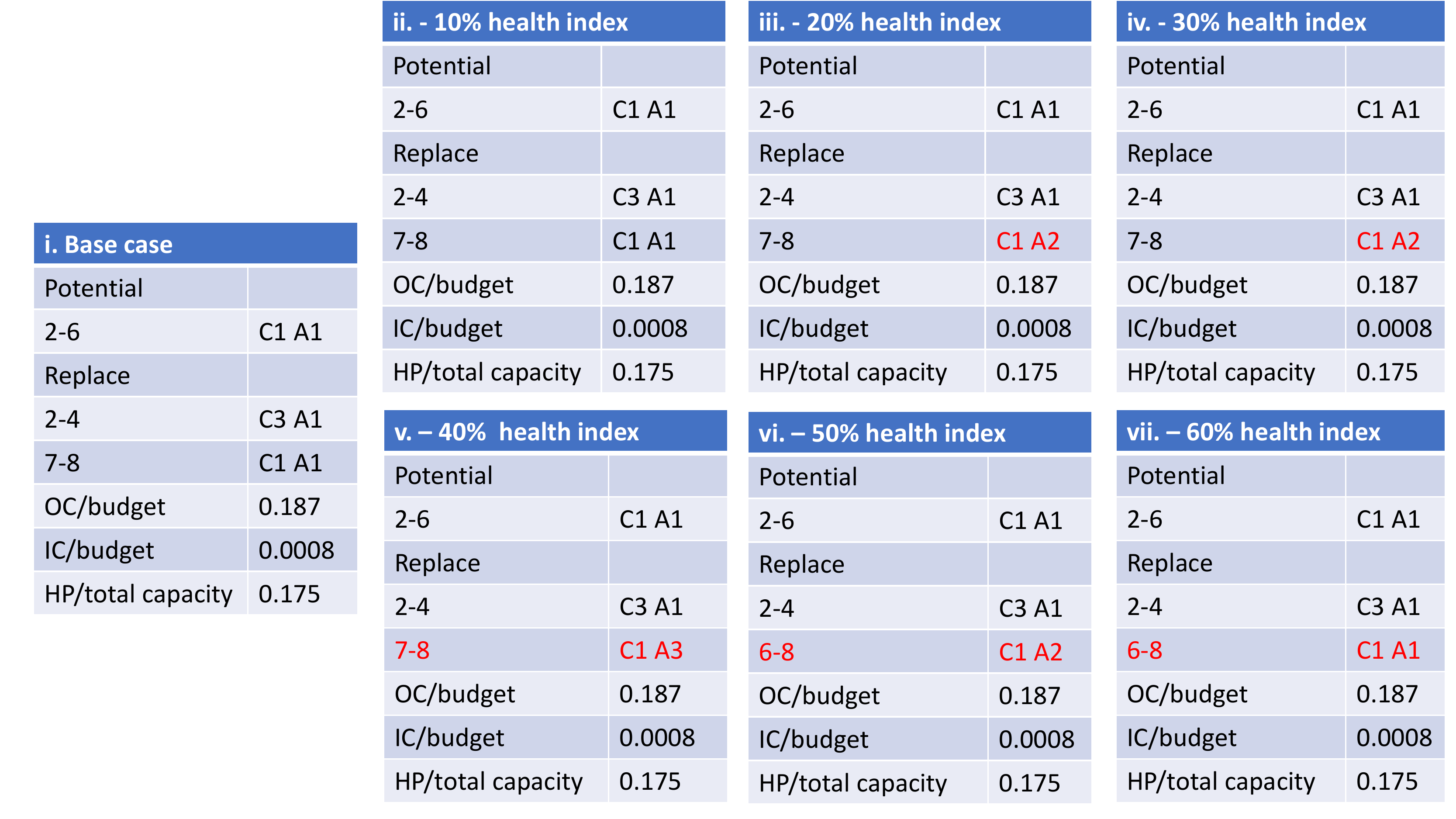}
    \caption{Decisions as the health index of tower 8  varies}
    \label{fig:s5c3}
  \end{subfigure}
  \begin{subfigure}{\textwidth}
    \includegraphics[width=\textwidth]{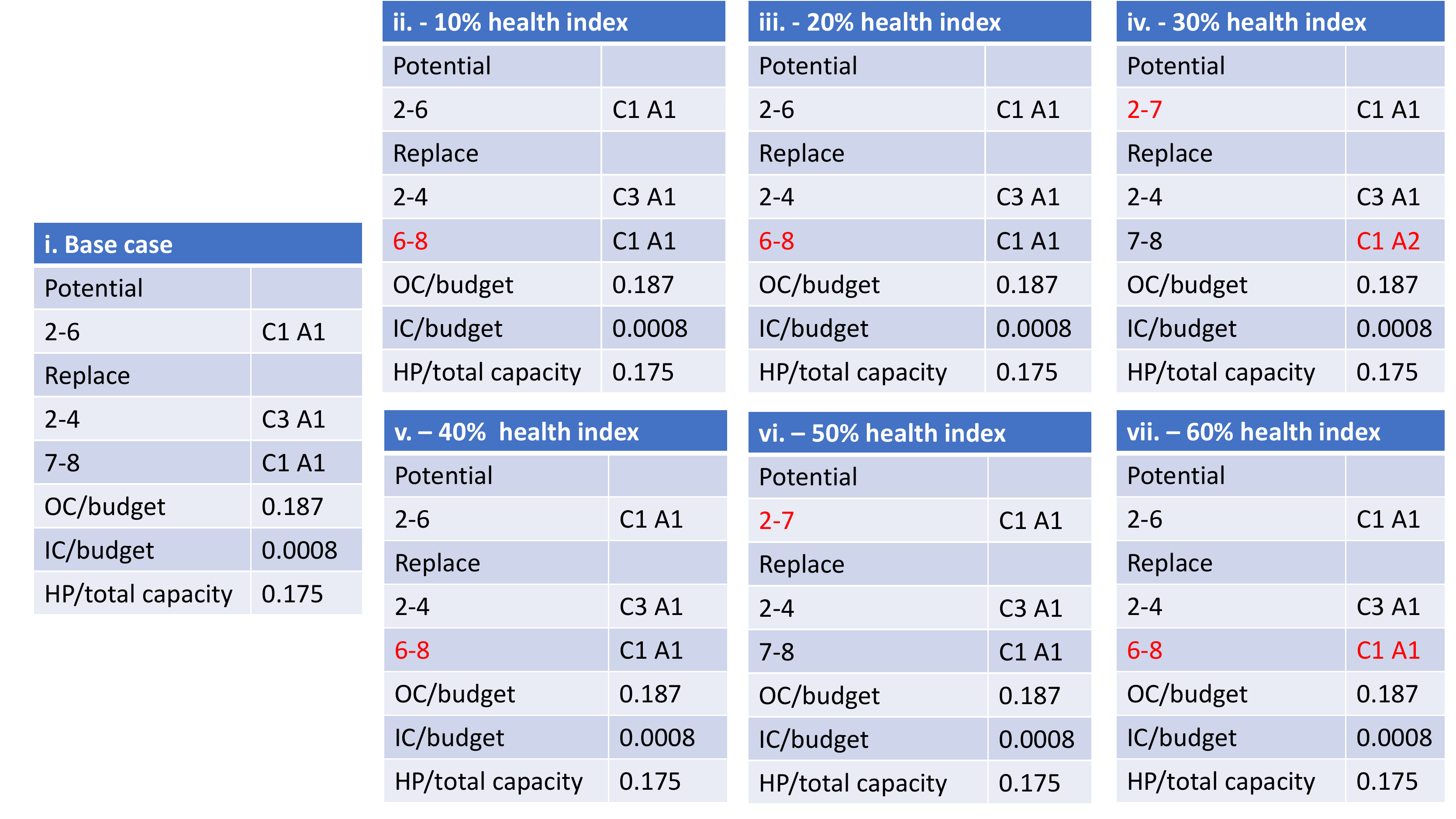}
    \caption{Decisions with change in the tower 2 \& 8 health index}
    \label{fig:s5c4}
  \end{subfigure}
     \caption{Sensitivity of tower 8, (2, 8) health indices}
    \label{fig:s5-2}
\end{figure}
\begin{figure}[H]
  \begin{subfigure}{\textwidth}
    \includegraphics[width=\textwidth]{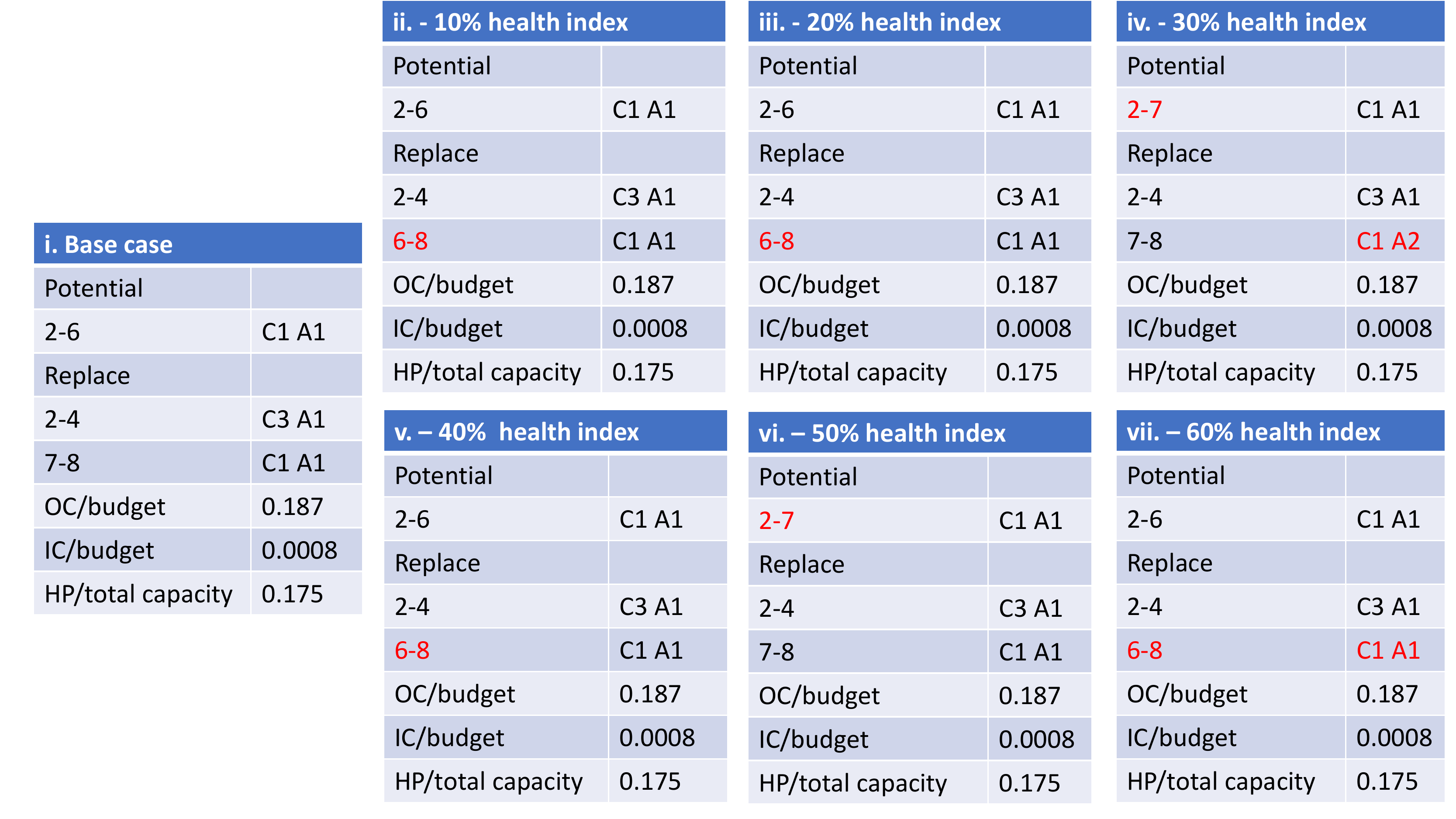}
    \caption{Decisions with change in the tower 4 health changing}
    \label{fig:s5c5}
  \end{subfigure}
  \begin{subfigure}{\textwidth}
    \includegraphics[width=\textwidth]{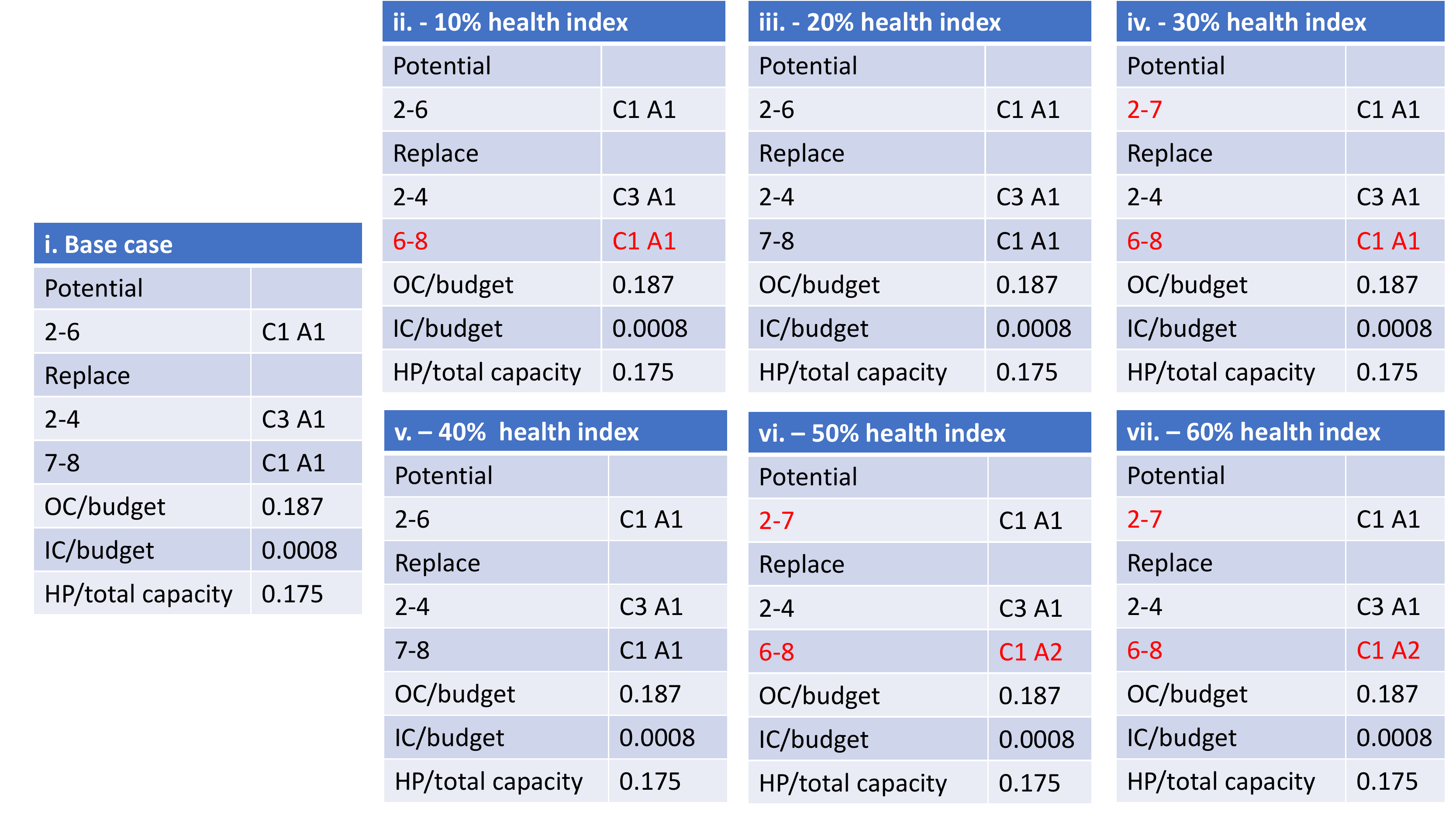}
    \caption{Decisions with change in the tower 2, 4, \& 8 health changing}
    \label{fig:s5c6}
  \end{subfigure}
    \caption{Sensitivity of tower 4, (2, 4, 8) health indices}
    \label{fig:s5-3}
\end{figure}

In fig. \ref{fig:s5-1}, \ref{fig:s5-2}, and \ref{fig:s5-3} the decisions and network interventions under various health indices are presented. Fig. \ref{fig:s5c1} presents the scenario when the heat energy consumption is rising and when the HI of tower 2 is half compared to the cases presented in previous sections. Fig. \ref{fig:s5c2} depicts the decisions when the condition of the tower 2 is changing. The sensitivity case with change in condition of tower 8, which is at the end of network, is presented in Fig. \ref{fig:s5c3}. Let's now move to the scenario where two towers are deteriorating simultaneously. The decisions under these conditions are detailed in Fig. \ref{fig:s5c4}. Decisions with degradation of tower 4 are displayed in fig. \ref{fig:s5c5}. Finally a critical network condition is put under investigation wherein towers 2, 4, 8 are deteriorating simultaneously. The results from this condition are presented in Fig. \ref{fig:s5c6}.

It is observed that as the health of poles deteriorates, there are changes in the pattern of network interventions. However, minor changes in the network restructuring are observed when health indices of poles are considered. It was observed that such changes were becoming more substantial as the construction time increased. It can be observed that the sole deterioration of the health of a tower is not a sufficient condition for network restructuring and reconfiguration.


\begin{figure}[!htbp]
  \begin{subfigure}{\textwidth}
    \includegraphics[width=\textwidth]{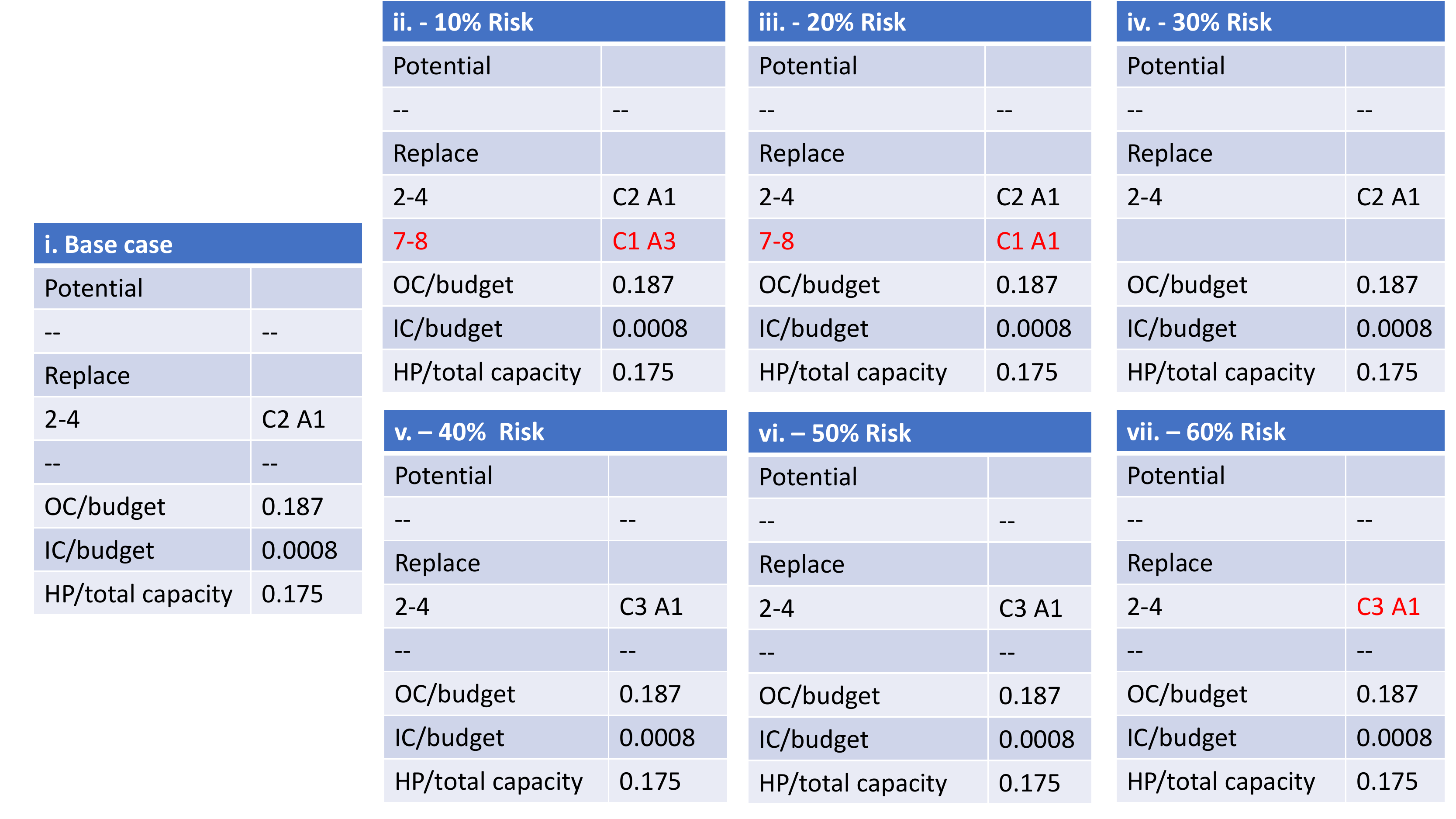}
    \caption{Sensitivity of construction time and tower health indices of tower 2,4}
    \label{fig:risk}
  \end{subfigure}
   \begin{subfigure}{\textwidth}
    \includegraphics[width=\textwidth]{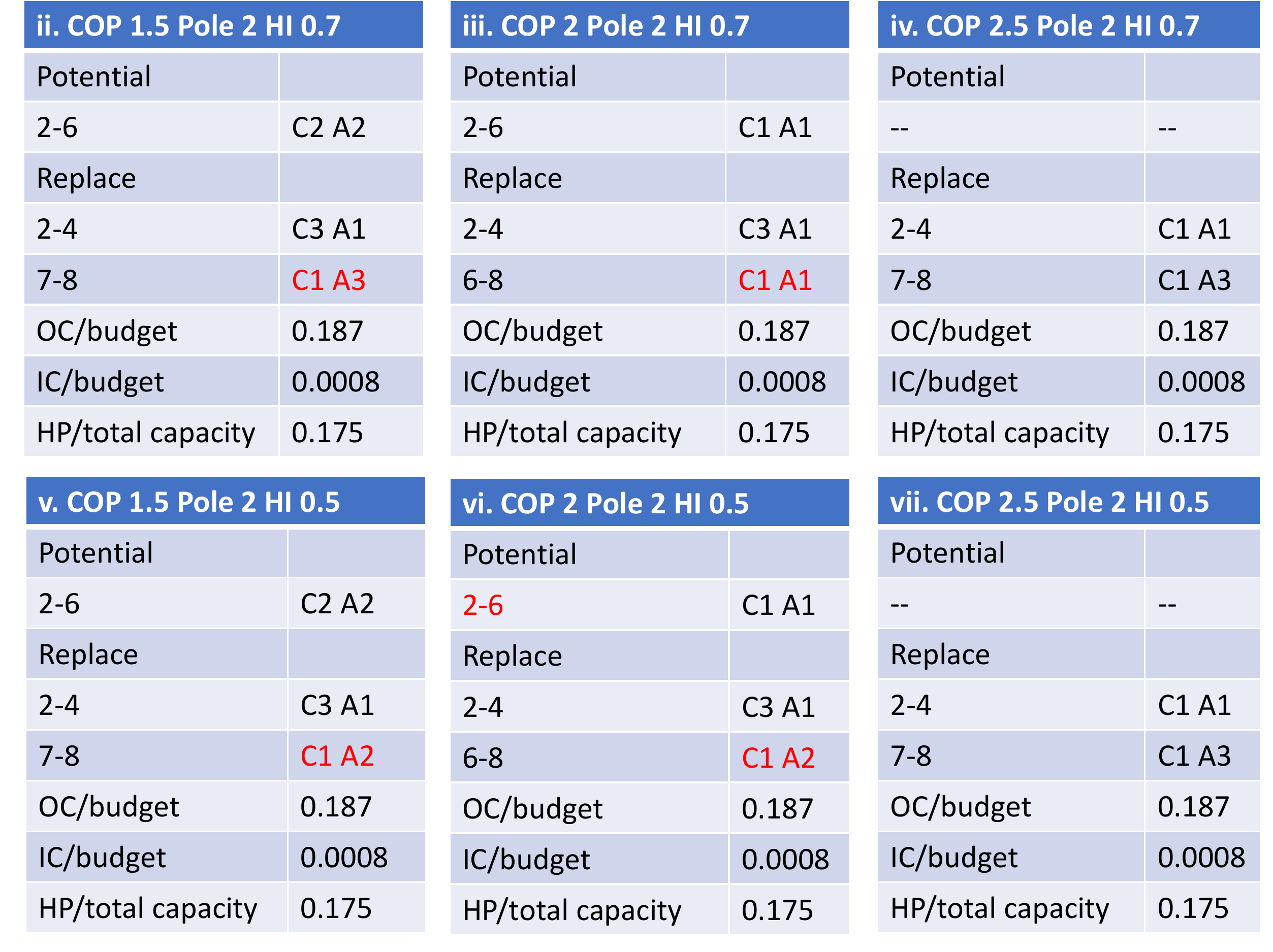}
    \caption{Sensitivity of construction time and tower health indices of tower 8}
    \label{fig:coppole}
  \end{subfigure}
    \caption{Decisions when the construction time, tower health are considered}
\end{figure}

\begin{figure}[!htbp]
  \begin{subfigure}{\textwidth}
    \includegraphics[width=\textwidth]{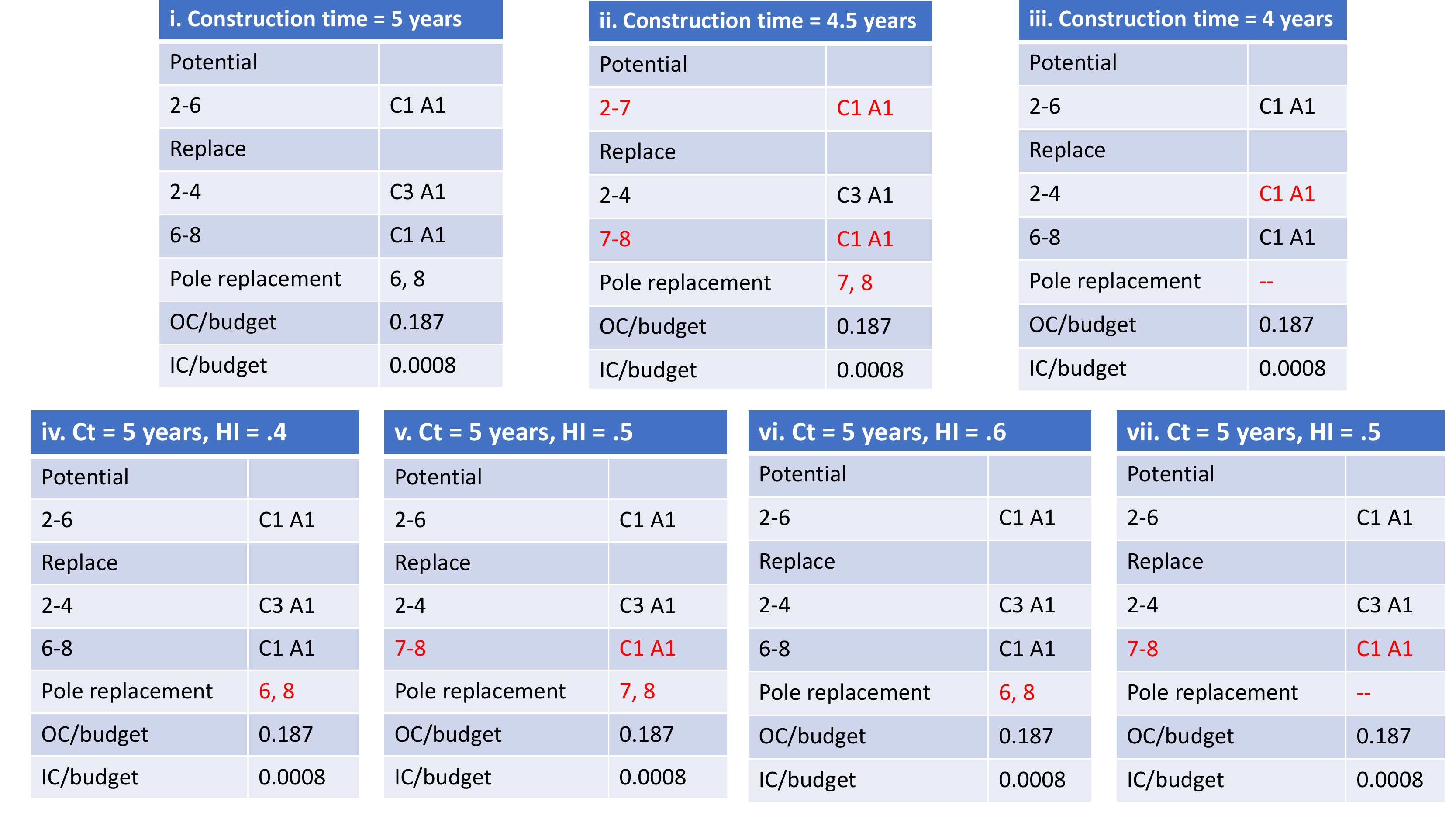}
    \caption{Sensitivity of construction time and tower health indices of tower 2, 4, 8}
    \label{fig:s6c1}
  \end{subfigure}
      \begin{subfigure}{\textwidth}
    \includegraphics[width=\textwidth]{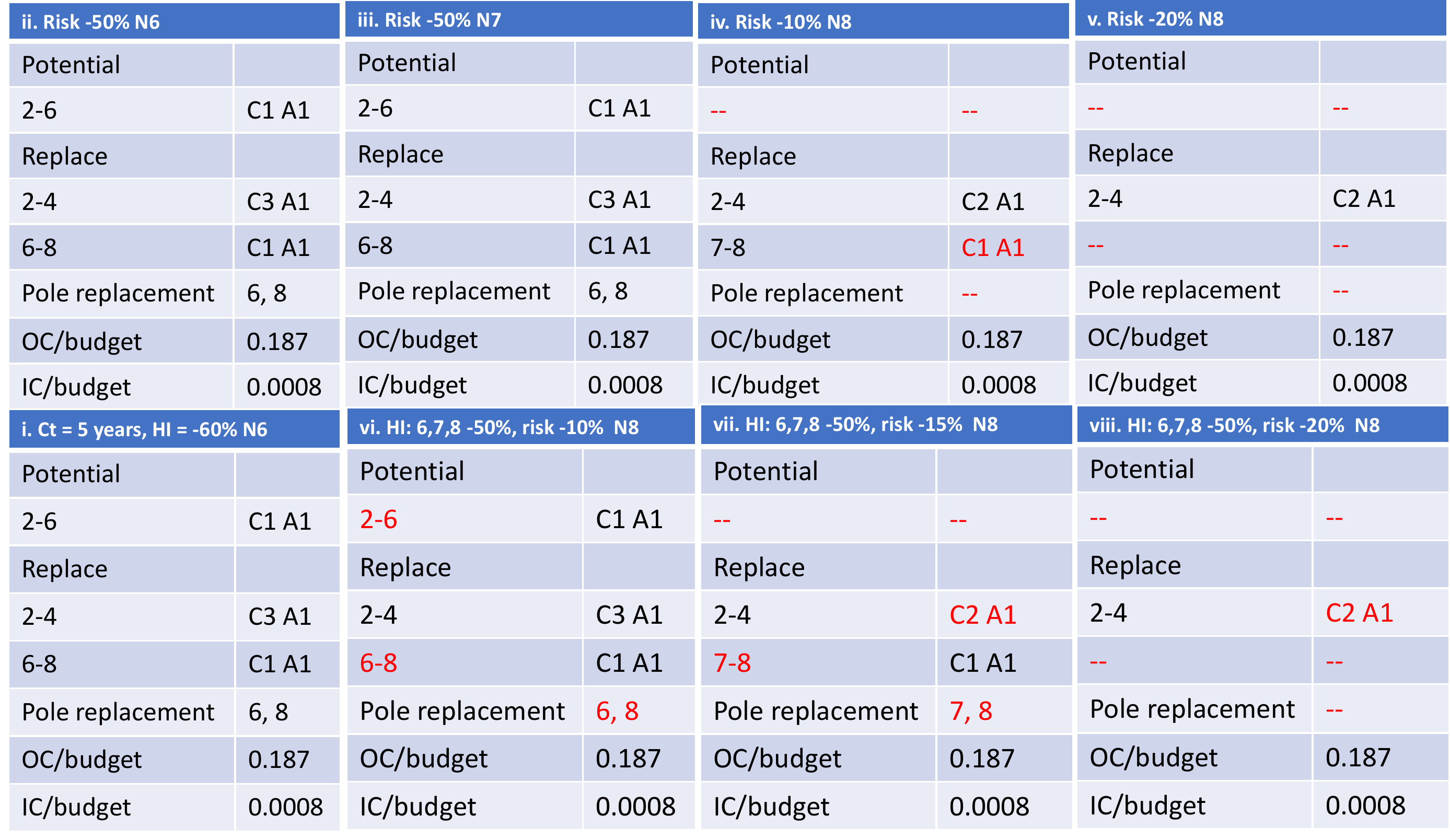}
    \caption{Decisions when risk is factored in}
    \label{fig:s6c2}
  \end{subfigure}
      \caption{Decisions when the construction time, tower health, and associated risk are considered}
\end{figure}

The construction of new lines or modifications of existing one could take 6 months to several years. According to the report from Berkeley lab on building electric transmission lines \cite{eto2016building}, there are a range of factors that contribute to risk in capacity building projects. Some of these factors are acquiring required permissions, funding acquisition, potential demand growth, and risks associated with civil construction works. In this work, a cumulative risk factor is used to represent a range of risks as listed before. Fig. \ref{fig:s6c2} presents the decisions when the construction time and risk factor are factored into the decision making process. Note that a particular node might have a lower health index indicating a lower operational capacity while having a lower risk. This translates to a scenario of important but not critical conditions. In this case, a load shedding is permitted to the same extent as the risk factor. In another scenario, wherein the risk is high and HI is low, all the demand must be met. Often such scenarios appears at sub-urban/rural areas at the edge of network with a week demand potential. Alternatively, an old line corridor in the urban area or a line connecting to an industrial consumer. 
Fig. \ref{fig:risk} presents the decisions when the risk associated to towers 2, 4 and 8 varies. Note that the the risk factor is 1 when the demand must be met at all point in time, while it reduces the flexibility through scope for load shedding increases. When the network operator deems the aforementioned towers to have less risky then the network restructuring decisions change. Particularly, with 10\% reduction in risk there is a new restructuring decision (line 7-8 with cable type 1 on year 3) in comparison to the base case (without any risk). It can be observed that as the risk decreases then the number of network interventions significantly decrease. When the risk factor decreases from 30\% to 40\% for instance, the optimal decision is to dismantle the existing line 2-4 on year 1. This indicates that the risk factor enables the network operator to take optimal decisions apt to the existing circumstances. Fig. \ref{fig:coppole} presents two set of scenarios where the health index of a tower and the COP of heat pump varies. This scenario sheds light into how the level of health of a tower could affect the network expansion. For example, when the COP is high (2.5) but the health is low (0.7), then there are no network reconfiguration or potential lines. while the COP is lower (2) then the optimal decision is to invest in new lines and perform restructuring. This illustrates that the network condition is an important factor in determining the minimum efficiency level of heat pump to be installed. This is particularly important in setting up the minimum parameters for the assets to listed in VPP. 

When the construction time was varied from 5 to 4 the decision in terms of network interventions and capacity building changed as presented in Fig. \ref{fig:s6c1}. The optimal decision, when the construction time is 5 years, is to replace poles (6 and 8), build new line capacity (2-6 using line C1 on year 1), replace existing line (6-8 using line C1 on year 1) and restructure the network by dismantling the line 2-4. The pole replacement is essentially a trade-off between the construction time and cost of replacement. In this case, if the construction time is longer than 4.5 years then the optimal decision is to replace the towers with a fewer network interventions. 
When the construction time reduces, the number and type of network interventions reduce significantly. Factoring the condition of the tower into the decision making process changes the overall network structure. In particular when the construction time is 5 years and condition of tower is 60\% of its original capacity, an optimal decision is to withheld from tower replacement. Similar to an urban power network, the risks at the edge of network (path 6-7-8) are now included into the decision making process. The results presented in Fig. \ref{fig:s6c2} illustrate how the risk impacts the optimal network expansion decision. When the risk in node 8 is 10\% then there are fewer network interventions as compared to the 50\% risk in node 7. Fixing the health of node 8 at 60\%, when the risk is 100\% then the optimal decision is to replace the towers 6 and 8. While the risk is reduced by 10\% then the tower replacements are not required. This implies that the health of the tower 8 is both critical and urgent given the risk. However, the health of tower 7 might be critical but the associated risk is low therefore not urgent. 



\begin{figure}[!htbp]
    \includegraphics[width=\textwidth]{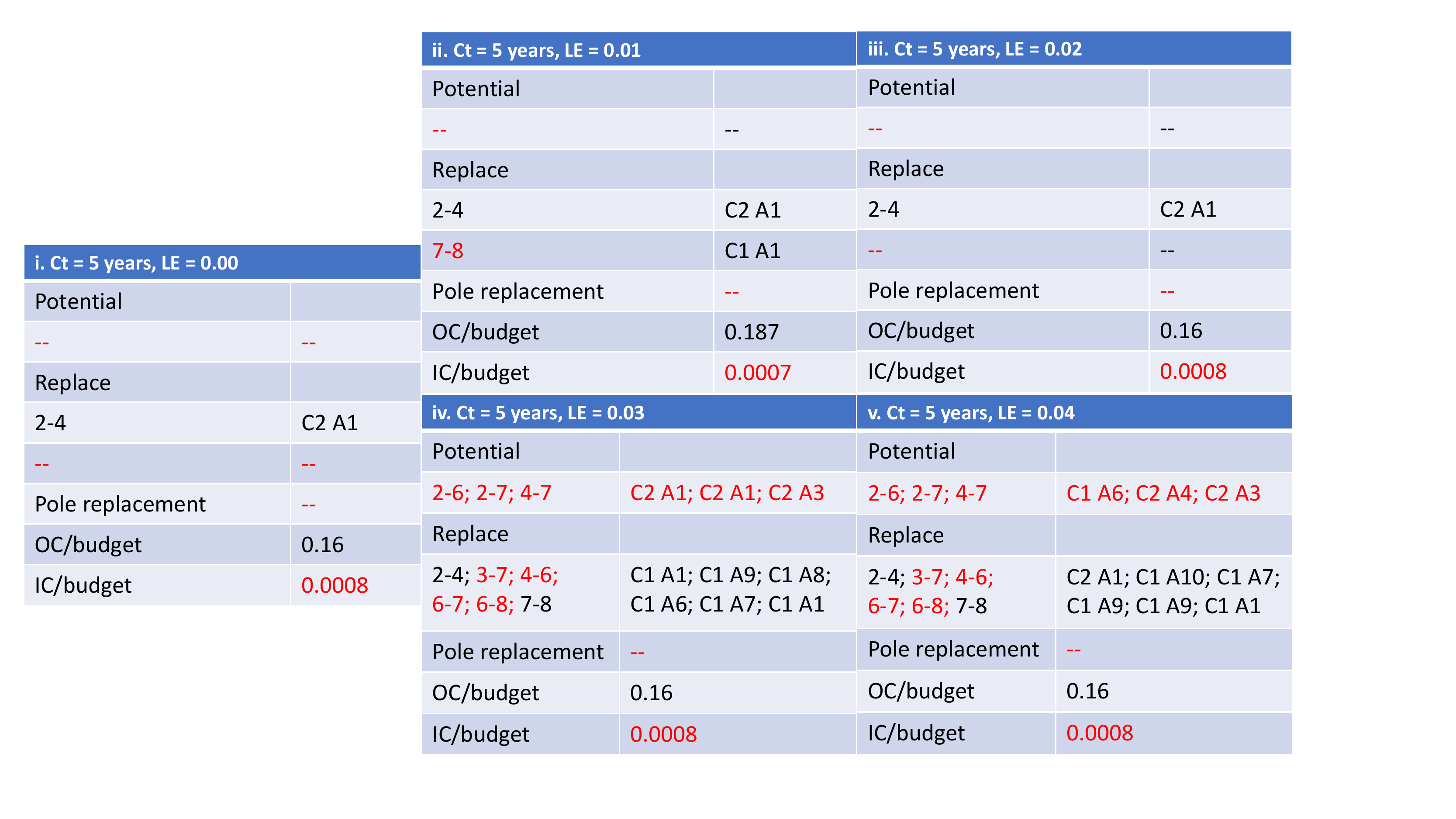}
    \caption{Variation in decisions with introduction of learning effect (LE)}
    \label{fig:s6c2_LE}
\end{figure}

Learning effect is the influence of experience gained over time of perming repetitive activities. In mathematical optimization this is realized through discounting the cost in the objective function. The authors have thoroughly studied and illustrated the effect of technological learning  in a previous work in \cite{bordin2021multihorizon}. In this paper, a sensitivity analysis is conducted to understand the impact of learning effect on the model decisions when the network condition is factored into the model parameters. The learning effect is gradually increased by 10\% and the results are detailed in the tables in fig. \ref{fig:s6c2_LE}. It can be observed from the results that the number of network interventions significantly increase as the learning effect increases. The reasoning behind the decisions is the effect of discounted cost. In particular, when the learning effect has increased by 30\% then a significant numbers of network interventions including reconfiguration decisions are observed. This is a pivotal point as further increments, from 30\% to 40\%, the changes in decision are trivial. 

\begin{figure}[H]
    \includegraphics[width=\textwidth]{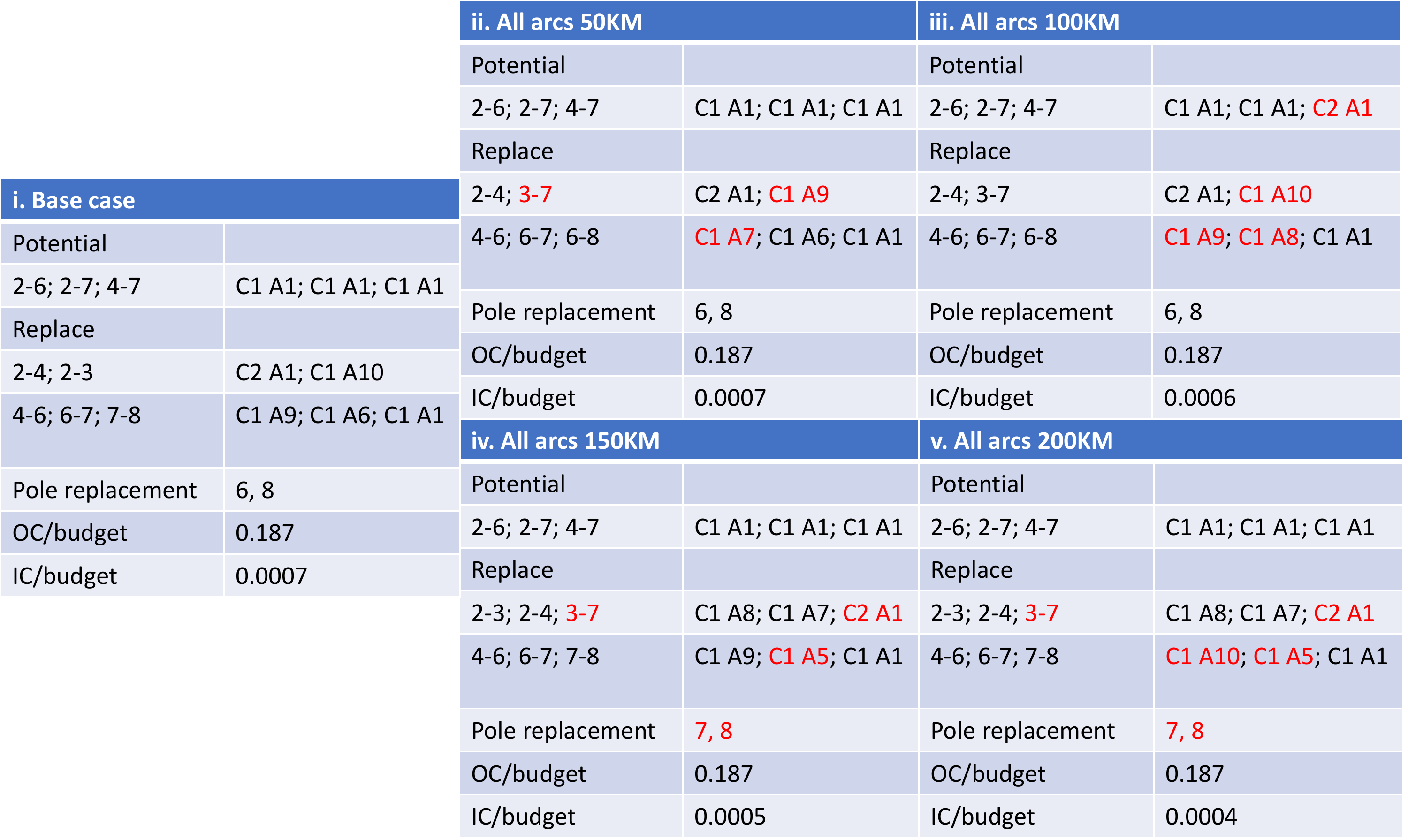}
    \caption{Decisions when the line lengths are taken into account}
    \label{fig:s63}
\end{figure}

Typically as the length of line to be constructed increases, the associated cost follows the same trend. Often in expansion planning models the actual length of lines is aggregated to simplify the computational burden. 
Next, how the length of existing lines impact the expansion decisions is investigated. The length of existing lines are increased by 10\% and the decisions are compared to a base case (with aggregated line lengths). The results are presented in the tables in fig. \ref{fig:s6c2_LE}. There is a significant change in terms of cost and network interventions as the line lengths are considered. As the line lengths start to increase, the number of interventions follows through and the main choice is shift towards a cheaper cable type. Since the cheaper cable has also a lower performance (i.e. lower capacity), the model has to include more interventions to get a mix that can allow the power flow and overcome the lower performance of cheaper cables. To illustrate in table \ref{fig:s63}.v has six replacements as part of network restructuring compared to table \ref{fig:s63}.i decisions. The location of demand and generation units and line capacities are among the key factors for this decision change. Thereby in an expansion decision making process the network condition plays a substantial role.   










\section{Conclusion and future scope}


This paper investigates the impact of resilient power network planning considering the condition of the network on the investment and operational costs for a virtual power plant in an integrated energy system. A virtual power plant provides a platform to facilitate local energy transactions with an integrated energy system. The proposed model takes into account both heating and electric demand. The choice to install local heat pumps to meet the heating energy consumption is among the decisions of capacity building. 
Reconfiguration, restructuring and tower replacement decisions are part of the network intervention options. The condition of network in the form of health index of tower, risk factor, maintenance costs of lines are part of the network information.A time horizon of 10 years in taken into account following the typical timeline for network reinforcement. A multi-horizon decision making process is followed where a decision can be taken within the set time horizon based on the current status and future projections. In addition to that, the effect of learning from experience is factored in to the decision making represented in form of learning effect. A wide range of sensitivity experiments are conducted to identify the trade-off between the network condition and capacity building. 


It is evident from the experimental results that the when the network condition is factored in, then the model decisions significantly vary than the scenario without network condition. Particularly the operational cost and network interventions. Inclusion of replacement decisions of towers along with lines reflect the practical reality of network operations. Moving forward, the level of importance of a tower through a risk factor demonstrates how the consumer type could influence the decision. Indeed in results it is clear that it is cost optimal to overlook condition of a tower if it is serving relatively low volume of growth.
In summary, inclusion of network condition to the decision process for network expansion planning demonstrates how the virtual power plants and utility operators can coordinate the decision making process. The allocation of risk factor is an effective way for the network owner to regulate the network interventions and reduce investments. 
In a future avenue various network structures could be investigated. The health index could be extended to also power lines to factor the changes on the lines. More critical analysis can be conducted in formulating the health index. 

\section*{Acknowledgement}
This work is supported by the Estonian Research Council grant PUTJD915.

\bibliographystyle{sec/helpers/elsarticle-num-names}
\bibliography{references.bib}

\end{document}